\documentclass[aps,prb,showpacs,raggedbottom,nobalancelastpage,amssymb]{revtex4}
\usepackage{amsmath}
\usepackage{amsfonts}
\usepackage{graphicx}
\usepackage{appendix}
\usepackage{dsfont}
\usepackage{amssymb}
\usepackage{bm}
\usepackage{color}
\usepackage{ulem}
\usepackage{soul}
\usepackage{natbib}
\usepackage{nccmath}
\usepackage{array}

\newcommand{\beq}{\begin{equation}}
\newcommand{\eneq}{\end{equation}}
\newcommand{\be}{\begin{equation}}
\newcommand{\ee}{\end{equation}}
\newcommand{\bea}{\begin{eqnarray}}
\newcommand{\eea}{\end{eqnarray}}

\usepackage{multirow}
\usepackage{wasysym}


\begin{document}

\title{Persistent current and zero-energy Majorana modes in a p-wave disordered superconducting 
ring }
\author{Andrea Nava$^{1}$, Rosa Giuliano$^{2}$, Gabriele Campagnano$^{3}$, and  
Domenico Giuliano$^{2}$}
 \affiliation{$^{1}$  Scuola Superiore di Studi Avanzati (SISSA),  Via Bonomea 265, I-34136 
Trieste, Italy\\
$^{2}$ Dipartimento di Fisica, Universit\`a della Calabria Arcavacata di Rende I-87036, Cosenza, Italy
and
I.N.F.N., Gruppo collegato di Cosenza, Arcavacata di Rende I-87036, Cosenza, Italy\\
$^{3}$ CNR-SPIN, Monte S. Angelo-Via Cintia, I-80126, Napoli, Italy and 
Dipartimento di Fisica ``E. Pancini" , Universit\`a di Napoli ``Federico II'', 
Monte S. Angelo-Via Cintia, I-80126 Napoli, Italy}
\date{\today}

\begin{abstract}

We discuss the emergence of zero-energy Majorana modes in
a disordered finite-length p-wave one-dimensional superconducting 
ring, pierced by a magnetic flux $\Phi$ tuned at an appropriate value 
$\Phi = \Phi_*$. In the absence of fermion parity conservation, 
we evidence the emergence of the Majorana modes by looking at 
the discontinuities in the persistent current $ I [ \Phi ]$ at 
$\Phi = \Phi_*$. By monitoring the discontinuities in $ I [ \Phi ]$, 
we map out the region in parameter space characterized by the 
emergence of Majorana modes in the disordered ring.
\end{abstract}


\maketitle
\section{Introduction}
\label{intro}

Majorana fermions,   particles coinciding with their own antiparticles, 
were proposed by Majorana in 1937 \cite{majorana}. While, so far, they 
have never been detected in a particle physics experiment
in the last years, after Kitaev's proposal that Majorana fermions may appear as 
zero-energy real fermionic modes [''Majorana modes'' - MMs] localized at the interface between a
p-wave one-dimensional superconductor and a normal metal \cite{kita}, 
the search for Majorana fermions in such systems has become one of the most 
relevant and promising areas in condensed matter physics \cite{wilczek}.

Besides Kitaev's proposal, the emergence of MMs in condensed matter systems has been predicted 
in a number of systems such as superconductor-topological insulator interfaces 
\cite{fukane,fukane_x,colin,beenakker_letter_add}, 
in proximity-induced superconducting quantum wires with strong spin-orbit interaction \cite{lutchyn,oreg_0,jelena,jelena2},
in helical magnets \cite{helimag}, in ferromagnetic atoms in proximity to superconductors 
\cite{choy,nady}. In this context, interesting phases with unconventional properties have been 
predicted at junctions between topological superconductors, hosting MMs at their endpoints, and interacting 
one-dimensional electronic systems (Luttinger liquids) \cite{alicef,giuaf_junction,giu_af1}. 
In addition, MMs emerging at junctions of one-dimensional interacting quantum wires \cite{beri_1,beri_2,beri_3,nava_1}, or of
systems formally described as interacting electronic systems, such as  quantum Ising spin chains 
\cite{tsvelik_Ising,tsve_2,tsve_3,offcrit}, one-dimensional XX- \cite{crampettoni}, or XY- \cite{gstt} models, or 
pertinently designed Josephson junction networks \cite{Giuso_x1}, have been predicted to give rise to the so-called 
''Topological Kondo Effect'', a remarkable realization of Kondo Effect in which the impurity spin, 
determined by the MMs, is nonlocal in the wire indices and, thus, cannot be expressed as a functional
of local operators \cite{beri_1,beri_2}. Finally, it is worth mentioning that, besides 
being of remarkable interest for fundamental physics, MMs are also 
of great interest for quantum computation since, due to their nonabelian statistics \cite{nayak}, 
they appear to be among the most natural candidates to work as robust qubits \cite{ivanov}. 

The proliferation of theoretical literature about Majorana fermions in condensed matter
systems has triggered a number of experimental attempts to probe MMs in pertinently 
designed devices. The main route followed in the experiments consists in measuring 
the effects in the transport across junctions between topological superconductors and 
normal metals possibly due to the presence of localized MMs at the interfaces 
\cite{exp_1,exp_2,exp_3}. Unfortunately, despite the excitement following early
 experimental results, the question of whether what is seen in a transport experiment
is actually due to the presence of a MM, or to other possible mechanisms, is still 
debated, with no ultimate answer so far given, mainly because of 
the high uncertainty about the possible physical processes taking place in 
the systems when it is connected to the metallic contacts required for  a
transport experiment \cite{debate_1,debate_2}. 
It becomes therefore crucial to engineer systems in which
the MMs can  be detected in noninvasive experiments, different from a transport 
measurement. In this direction, an interesting proposal has been put forward in 
 Ref.[\onlinecite{procolo}], where it was proposed to realize MMs in a frustrated, 
 finite size topological superconducting quantum interference devide (SQUID) at pertinent values of the applied magnetic 
flux piercing the superconducting ring, as well as in 
Refs.[\onlinecite{zha_1,zha_2}], where the Majorana zero mode
and the persistent spin current are investigated in 
mesoscopic d-wave-superconducting loops in the 
presence of spin-orbit  interaction and in mesoscopic s-wave superconducting loops. 
The advantage of such  proposals is twofold: on one hand, it implies the possibility of recovering 
MMs in a finite system, once the applied flux is properly tuned (differently from  
what happens, for instance, in the Kitaev model, where, rigorously speaking, 
''true'' zero-energy MMs are in general recovered either in the infinite chain limit, or as a 
result of a challenging fine-tuning of the system parameters \cite{kita}); on the other hand, 
it provides an example of MMs realized in a controlled way in a system with tunable 
control parameters, such as a quasi one-dimensional SQUID ring. 

In fact, on the theoretical side, in the last years 
the study of  quasi one-dimensional superconducting rings has been
taken advantage of the systematic application of effective field theory approaches
  \cite{giu_sox01,giu_sox02}, which allowed to study nontrivial effects arising in 
pertinently designed one-dimensional superconducting devices, such as frustration of 
decoherence \cite{giu_sox2,cirillo}, correlated 
hopping of pairs of Cooper pairs \cite{giu_sox3,giu_sox4},  etc. On the experimental 
side, the  recent progresses  made in the fabrication of nanostructures and, in particular, 
of superconducting and/or hybrid rings,   where superconductivity is induced by
proximity effect  only in part of the ring \cite{rings},  provides 
an excellent level of control on the design and fabrication of systems 
which are likely to host MMs, at pertinent values of their parameters.

Yet, notwithstanding the good control one may achieve on the fabrication parameters
of the system, the unavoidable presence  of disorder can still affect the final behavior 
of the device. In fact,  disorder is in general  known to have drastic consequences
for the properties of one dimensional normal electronic \cite{Beenakker_review} and 
superconducting systems \cite{been_2014}. Therefore, it is  important
to clearly spell out the stability of MMs against disorder. For instance, an open, finite-length
Kitaev chain Hamiltonian breaks  both  spin-rotational and time-reversal invariance.
Therefore, it fall into class D of  the  Altland-Zirnbauer ''tenfold way'' classification of 
disordered systems in relation to the symmetries of the corresponding 
Hamiltonian \cite{alt_zirn,been_2014}. When the chain is 
within its  topological phase, characterized by the presence of MMs, 
a weak disorder gives rise to rare small-size nontopological regions, embedded within 
the topological background. At any interface between topological and nontopological 
regions, additional MMs arise which, for small disorder, hybridize across the 
nontopological region into Dirac quasiparticle states that start to fill in the 
gap. On increasing the disorder strength, the nontopological regions 
start to proliferate. The corresponding proliferation of MMs and, correspondingly, 
of subgap states,  strongly renormalizes the density of states (DOS) inside the gap, 
eventually leading to low-energy singularities because of the Griffiths 
effect in the finite wire \cite{motrunich,sau_2,been_2014}. At strong 
disorder, the hybridization between MMs at the endpoints of the chain and   
zero-modes located at the interfaces between topological and nontopological regions 
eventually washes out the  former ones, thus  
driving the system across a disorder-induced topological phase transition
\cite{motrunich,sau_2,been_2014,brouwer_1,brouwer_2,pientka_1,pientka,shuricht,wish}.

In this paper, we discuss the emergence of MMs at a disordered finite-length p-wave one-dimensional
superconducting ring (PSR), pierced by a magnetic flux $\Phi$ [which, throughout the whole paper, 
we measure in units of the quantum of flux $\Phi_0^* = h c /(2e)$]. In the 
absence of disorder, due to the 
finite size of the system, in general one expects not to find ''true'' MMs at
zero energy, but rather two finite-energy subgap Dirac modes, due to 
the hybridizations between the  MMs through the finite length of 
the system, into two ''putative Majorana fermions'' (PMFs), whose energies $\pm \epsilon_0 [ \Phi ]$ 
are disposed symmetrically with respect to the Fermi level  \cite{kita}. Nevertheless, we show that, 
under quite generic conditions on the PSR parameters, in the absence of 
disorder it is always possible  to tune $\Phi$ at a value $\Phi_*$ (that is a function of the parameters of the ring) 
at which the subgap modes appear exactly at 
zero energy, due to the level crossing (LC) between the PMF energy levels, 
thus giving back two true zero-energy MMs.

To probe the PMF LC, we look at the dependence of the persistent current induced in 
the ring by the applied flux, $I [ \Phi ]$, with respect to $\Phi$. In the absence of 
fermion parity (FP) conservation (which is quite a natural assumption in a quasistatic 
process), at any PMF LC the PSR relaxes to the minimum energy states, thus 
triggering  a discontinuity  in $ I [ \Phi ]$ at $\Phi = \Phi_*$. Therefore, in a clean 
PSR, discontinuities in $ I [ \Phi]$ are uniquely associated to the emergence of 
zero-energy MMs. The key point of our paper is that this correspondence is 
preserved in the presence of  (a limited amount of) disorder. Specifically, 
on introducing disorder in the ring and looking at the discontinuities in
$ I [ \Phi ]$ at  $\Phi = \Phi_*$ we identify a region 
in the parameter space  (strength of disorder - chemical potential  plane), in which $ I [ \Phi ]$ is 
discontinuous at any realization of disorder. We therefore interpret this result as an evidence 
for  the emergence of MMs at the ring even in the  presence of disorder. According to 
this criterion, we map out   the corresponding region in the  strength of 
disorder - chemical potential phase plane, which we 
refer to as  ''putative topological phase'' (PTP).  We find that, while, on one hand, the PTP   derived within out 
method strongly resembles the one found for an infinite Kitaev chain by using transfer 
matrix (TM) approach \cite{shuricht,wish}, in our case the actual phase boundary appears 
broadened, as it must be as a consequence of Griffiths effect for the finite chain 
near by the phase transition  \cite{motrunich,sau_2}. Nevertheless, differently from 
the TM technique, our approach can be readily implemented as an actual experimental procedure: it 
is enough to map out the persistent current and to look at possible discontinuities as a 
function of the applied flux.

   We regard our result in the clean limit as a generalization of the derivation of 
Ref.[\onlinecite{procolo}] to a generic p-wave superconducting ring with 
a weak link. The crucial requirements for our approach to work are the 
absence of FP conservation and the existence of a LC at $\Phi = \Phi_*$, which leads
to a $2 \pi$-periodicity of $I [ \Phi ]$, with a discontinuity at $\Phi = \Phi_*$.
In this respect, despite the apparent similarity in the discussed device, our
approach is fully complementary to the one  
discussed in Ref.[\onlinecite{pientka}]. Indeed, to probe the emergence of subgap PMF
levels, in Ref.[\onlinecite{pientka}] a nonzero hybridization between the localized 
MMs is required to take place via the p-wave superconducting region. 
This assures the persistence of a $4 \pi$-periodic component of $ I [ \Phi ]$ even 
in the absence of FP conservation, which avoids the  LC at $\Phi_*$ just because of 
the opening of the hybridization gap. Accordingly, the main challenge in probing the 
PMFs in the presence of disorder just relies on detecting the 
corresponding survival of the $4 \pi$-harmonics in the current. Instead, we do assume that 
there is no hybridization between the MMs via the finite length of the p-wave superconductor.
This yields a $2 \pi$-periodicity of $I [ \Phi ]$, with a discontinuity at $\Phi = \Phi_*$
that, in our approach, becomes the main fingerprint of the emergence of PMFs, in the clean 
as well as in the disordered ring.  

The paper is organized as follows:

\begin{itemize}
 \item In Sec. \ref{model_hamiltonian} we present the system Hamiltonian for the PSR in the absence of 
 disorder, review the calculation of the persistent current and discuss the conditions under which 
 subgap PMFs undergo a LC at some value $\Phi_*$ of the flux $\Phi$. Eventually, we discuss the relation between PMF LCs and 
 discontinuities in $I [ \Phi ]$.
 
 \item In Sec. \ref{disorder_1}, we perform   a detailed analysis of the DOS in the presence 
 of disorder, with particular emphasis onto the subgap PMFs and on their dependence on  $\Phi$. We show
 that, provided the system parameters in the ''clean'' limit are 
chosen so that there is a PMF LC, the LC survives the presence of disorder so that either 
there are still PMF states and they cross at a pertinent value of $\Phi$, or PMF states are 
washed out by strong disorder and, accordingly, the zero-energy LC disappears.   

\item In Sec. \ref{PTP_boundaries},  we show how the relation between PMF LCs and 
 discontinuities in $I [ \Phi ]$ extends to the disordered PSR. Eventually, we use 
 this result  to map out the whole  PTP in the disorder strength-chemical potential plane.
 
 \item In Sec. \ref{concl}, we summarize and comment our results and provide possible further 
 developments of our work.
 
 \item In Appendix \ref{open_chain}, we review the derivation of the eigenvalues and of the eigenfunctions
 of the open Kitaev chain, with particular emphasis onto subgap states.
 
 \item In Appendix \ref{closed_ring}, we review the derivation of the eigenvalues and of the eigenfunctions
 of the PSR. In particular, we find under which conditions on the system parameters there is 
 a PMF LC at $\Phi = \Phi_*$ and find an exact formula for $\Phi_*$ in terms of the system parameters.

\end{itemize}

\section{Model Hamiltonian and subgap states}
\label{model_hamiltonian}
 
\begin{figure}
\includegraphics*[width=.5\linewidth]{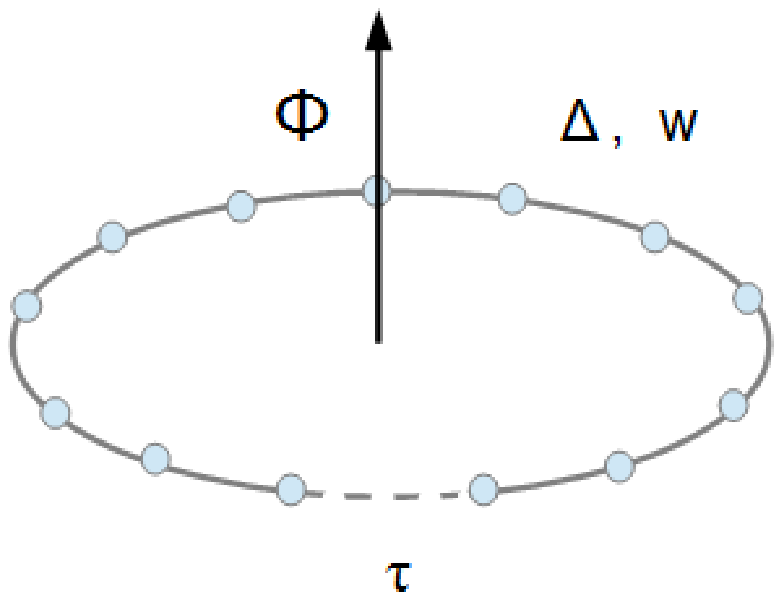}
\caption{Sketch of the one-dimensional p-wave superconducting ring pierced by 
a magnetic flux $\Phi$ described by $H_K$ in Eq.(\ref{mh.1}) plus 
$H_\tau [ \Phi ] $ in Eq.(\ref{mh.2}). } \label{system}
\end{figure}

In this section we discuss our system in the absence of disorder. In Fig.\ref{system} we 
provide a sketch of the PSR; it   is realized as a p-wave 
superconducting ring interrupted by a weak normal link and  pierced by a magnetic flux $\Phi$, which 
induces a persistent current $I [ \Phi ]$   through the system. To formally describe the 
p-wave superconductor we use Kitaev's one-dimensional lattice model Hamiltonian (KMH) \cite{kita}. 
KMH can be regarded as an effective low-energy description of a  quantum wire with  a  strong spin-orbit coupling and 
a large enough Zeeman effect, which turns into a one-dimensional p-wave superconductor by proximity to 
a ''standard'' s-wave bulk superconductor \cite{lutchyn,oreg_0}.

\noindent

The Kitaev lattice Hamiltonian for a one-dimensional p-wave superconductor is given by \cite{kita}

 \beq
H_K = - w \sum_{  j = 1}^{ \ell - 1} \{ c_j^\dagger c_{ j +1 } + c_{ j + 1}^\dagger c_j \} - \mu \sum_{ j = 1}^\ell c_j^\dagger c_j 
+ \Delta \sum_{ j = 1}^{ \ell - 1} \{ c_j c_{ j + 1} + c_{ j +1}^\dagger c_j^\dagger \}
\:\:\:\: .
\label{mh.1}
\eneq
\noindent
Following the notation of Ref.[\onlinecite{kita}], in 
Eq.(\ref{mh.1}) we denote  with  $c_j \: (c_j^\dagger )$ ($j = 1 , \ldots , \ell$) 
the  single-fermion annihilation (creation) operators
defined on site-$j$ of the one-dimensional chain, which  satisfy the canonical anticommutation relations 
$\{ c_j , c_{j'}^\dagger \} = \delta_{ j , j'}$, all the other anticommutators being equal to 0. 
We then denote with $w$ and $\Delta$  
respectively the normal single-electron hopping amplitude and the p-wave superconducting pairing, and 
with $\mu$ the  chemical potential. For the sake of simplicity, without any loss of generality, we further simplify 
$H_K$ by choosing   $  w = \Delta$  (which does not qualitatively affect the spectrum and the 
eigenfunctions with respect to the general case) and  $\mu \geq 0$ (the complementary situation $\mu < 0$
can be easily recovered by symmetry). Besides its mathematical simplicity, it is also worth 
noticing that the Hamiltonian in Eq.(\ref{mh.1}) with $w = \Delta$ takes a precise physical meaning, as it can 
be obtained from the Hamiltonian for an open quantum Ising chain via Jordan-Wigner transformation 
\cite{jordanwigner}.

As we review  in Appendix \ref{open_chain}, in its topological phase, 
a long open Kitaev chain  hosts zero-energy MMs localized at the endpoints of the chain \cite{kita}. 
The MMs can then be combined into a 
zero-energy Dirac mode, which implies a twofold spectral degeneracy of $H_K$, 
with degenerate eigenstates differing from each other by the total FP 
corresponding to the zero-energy mode being populated, or empty. For a finite-length chain (that is, 
with $\ell$ of the same order as  the superconducting coherence length of the p-wave superconductor, 
$\xi_0$), the MMs are hybridized by means of an overlap matrix element that is $\sim e^{ - \frac{\ell}{\xi_0}}$.
In this case, strictly speaking, MMs are not anymore true eigenstates of $H_K$. Instead, one may rather 
speak of two PMFs that hybridize into a finite-energy
Dirac mode, with corresponding disappearance of the fermion-parity related degeneracy. As we 
review in Appendix \ref{open_chain}, MMs as well as PMFs only emerge   provided 
$\frac{2 w}{ \mu  } >1$. Strictly speaking, one may dub such a phase ''topological'' only when MMs lie exactly at 
zero energy.  In general, we rather speak of a PTP, with 
corresponding PMFs hybridized into nonzero energy Dirac modes. Instead, neither MMs, or PMFs, appear in
the spectrum for  $\frac{2 w}{ \mu  } <  1$. To trade the open Kitaev chain for a PSR, we 
add to $H_K$    a normal weak link hopping term $H_\tau$. Defining $\tau$ to be the normal 
hopping amplitude and taking into account that    the applied flux $\Phi$  can be moved onto 
the weak link hopping term  by means of a simple canonical  transformation of the fermionic operators, 
$H_\tau$ can be presented as \cite{ngcg}

\beq
H_\tau [ \Phi ] = - \tau \{ e^{ \frac{i}{2} \Phi} c_1^\dagger c_\ell +  e^{ - \frac{i}{2} \Phi} c_\ell^\dagger c_1 \} 
\:\:\:\: . 
\label{mh.2}
\eneq
\noindent
In the following, we will use  the full Hamiltonian $H [ \Phi ] = H_K + H_\tau [ \Phi ] $ to compute the DOS  
and the persistent current $ I [ \Phi ]$. In general, at temperature $T$, the   persistent current  $I [ \Phi; T ]$  
is obtained from the free energy ${\cal F} [ \Phi ; T ]$ as (see, for instance, Ref.[\onlinecite{Beenakker_0}] for 
a review on this approach)

\beq
I [ \Phi ; T ] = e \partial_\Phi {\cal F} [ \Phi ; T ]
\;\;\;\; . 
\label{mh.7}
\eneq
\noindent
Throughout all our paper we will be focusing onto the   $T= 0 $-limit, in 
which Eq.(\ref{mh.7}) becomes 

\beq
I [ \Phi ] = e \partial_\Phi   E_{\rm GS}  [ \Phi  ]
\:\:\:\: , 
\label{mh.7bis}
\eneq
\noindent
with  $E_{\rm GS}  [ \Phi  ]$ being  
the total groundstate energy of the system. Therefore, in terms of 
the energies of the quasiparticle excitations of $H [ \Phi ]$, 
$\{ \epsilon_n [ \Phi ] \}$,  one  obtains 

\beq
I [ \Phi ] = e \partial_\Phi E_{\rm GS}  [ \Phi  ] =  e \partial_\Phi \sum_{\epsilon_n<0}
\epsilon_n [ \Phi ] 
\;\;\;\; , 
\label{mh.8}
\eneq
\noindent
with the sum  taken, as specified,  over negative-energy  single-quasiparticle states.
In a ring made with a conventional s-wave superconductor, as well as in a
 ring made with a p-wave superconductor in the nontopological phase, $ I [ \Phi ]$ is typically 
a periodic function of $\Phi$ with period $2 \pi$. Instead, when the p-wave superconductor 
is within its topological phase, if FP is conserved, 
the presence of MMs at the endpoints of the superconductor (and, therefore,
at the two sides of the weak link) typically makes $ I [ \Phi ]$ periodic with 
period equal to $4 \pi$  \cite{been_x1} (note that, for a finite size ring one should 
carefully spell out whether the $4 \pi$-periodicity is really due to FP conservation
in the presence of MMs, or is a simple effect of the crossover from a superconducting to 
a mesoscopic ring as $\ell \leq \xi_0$ \cite{montam}).  
The    $4 \pi$-periodicity simply derives from the fact that FP conservation forbids    relaxations 
  between the two levels $\pm \epsilon_0 [ \Phi ]$  
\cite{pientka}. This leads to  the so-called ''fractional Josephson effect'' (FJE), 
which has been proposed as an effective way  for evidencing the existence of  MMs \cite{been_x1}.
In fact, in reality it is  quite difficult to avoid FP non-conserving relaxation 
processes in quasistatic (DC) measurements \cite{pientka,law_lee_ng} and,  to overcome 
such a limitation, proposals of detecting FJE in nonstatic 
settings have been put forward in measurements of e.g. AC Josephson effect 
\cite{ac_1,ac_2}, finite-frequency current noise \cite{finite_frequency}, and 
Shapiro steps \cite{shapiro}.

In general, in a PSR,  one  expects a coexistence of a $2 \pi$- and of a $4 \pi$-harmonics, 
due to the existence of 
two possible ''channels'' for the MMs to hybridize into a Dirac mode:  
through the finite-length p-wave  superconducting chain, as well as via the weak link. 
The presence of the two harmonics and the value of 
their relative weight can be readily understood from our equation for the 
energy eigenvalues of the PMFs, Eq.(\ref{sg.38}) of Appendix \ref{closed_ring}, 
and from the exact condition for recovering a PMF zero-energy level crossing  
at pertinent values of $\Phi$,  Eq.(\ref{sg.47}) of 
appendix \ref{closed_ring}. Indeed, from Eq.(\ref{sg.38}) one  expects 
that $E_{\rm GS} [ \Phi] $ and, correspondingly, $ I [ \Phi ]$ 
are  $4 \pi$-periodic functions of $\Phi$, as they 
only depend on periodic functions of   $\Phi / 2$. 
As this picture strongly relies upon assuming FP conservation, which is 
hard to recover in a quasistatic DC current measurement, in Ref.[\onlinecite{pientka}]
it was noted that, since FP  changing processes are expected to take place 
via relaxation processes  that happen just at the PMF LCs,  one may just get rid of  
LCs by pertinently tuning the system parameters. Formally, we  rigorously 
state it in Eq.(\ref{sg.47}) of appendix \ref{closed_ring}, where, 
we prove that PMF LC takes place at $\Phi = \Phi_*$, with 
$\Phi_*$ satisfying the equation

\beq
\frac{2 w \tau}{\mu^2 - \tau^2 } \cos \left ( \frac{\Phi_*}{2} \right) =   e^{  - ( \ell - 2 ) \delta_0} \Rightarrow
\cos \left( \frac{\Phi_*}{2} \right) = \frac{\mu^2 - \tau^2}{2 w \tau }  e^{  - ( \ell - 2 ) \delta_0}
\;\;\;\; , 
\label{mh.3}
\eneq
\noindent
with $\delta_0$ defined as 

\beq
\delta_0 = 2 \sinh^{-1} \left\{ \sqrt{\frac{\Delta_w^2  }{8 w \mu}}
\right\}
\;\;\;\; .
\label{mh.4}
\eneq
\noindent
From Eq.(\ref{mh.3})   we see that 
a PMF LC  happens whenever the contribution to the splitting energy associated to the MM-hybridization
through the finite superconducting chain (that is, the term $\propto e^{ - \ell \delta_0 }$)
becomes equal, though opposite in sign, to the contribution  associated to MM-hybridization through 
the weak link (which is $\propto \tau$). Clearly, as the latter contribution is modulated by 
$\cos \left( \frac{\Phi}{2} \right)$, the level crossing can only happen provided   one recovers the condition

\beq
\left|  \frac{\mu^2 - \tau^2}{2 w \tau }  e^{  - ( \ell - 2 ) \delta_0} \right| \leq 1
\;\;\;\; ,
\label{condition}
\eneq
\noindent
that is, provided that either the chain is long enough (as $\delta_0 \sim \xi_0^{-1}$), or 
the coupling $\tau$  is strong enough, or both. Thus, to avoid PMF LC one has 
to make it impossible to satisfy Eq.(\ref{mh.3}) at 
any value of $\Phi$.   This strategy was actually pursued  in 
Ref.[\onlinecite{pientka}], where the regime complementary to ours
was recovered by assuming $\tau / w \ll 1$ 
and, at the same time, by considering the PSR close to the topological 
phase transition, at which $\xi_0 \to \infty$, so to 
make $\left|  \frac{\mu^2 - \tau^2}{2 w \tau }  e^{  - ( \ell - 2 ) \delta_0} \right| \gg 1$. 
In this case, the hybridization between the MMs through the finite p-wave superconducting
region yields a consistent spectral weight for the $4 \pi$-harmonics, which 
appears as a modulation of the $2 \pi$-periodicity in $I [ \Phi ]$,
 as a result of the competition between the 
''Kondo-like'' hybridization between the PMFs mediated by the 
weak link  \cite{giu_af1}, and the  ''RKKY-like'' interaction mediated by the finite 
chain length. When going across the phase transition,  the 
PMFs disappear, thus determining a full disappearance of the $4 \pi$-harmonics and
a purely $2 \pi$ periodic persistent current.   

\begin{figure}
\includegraphics*[width=1.\linewidth]{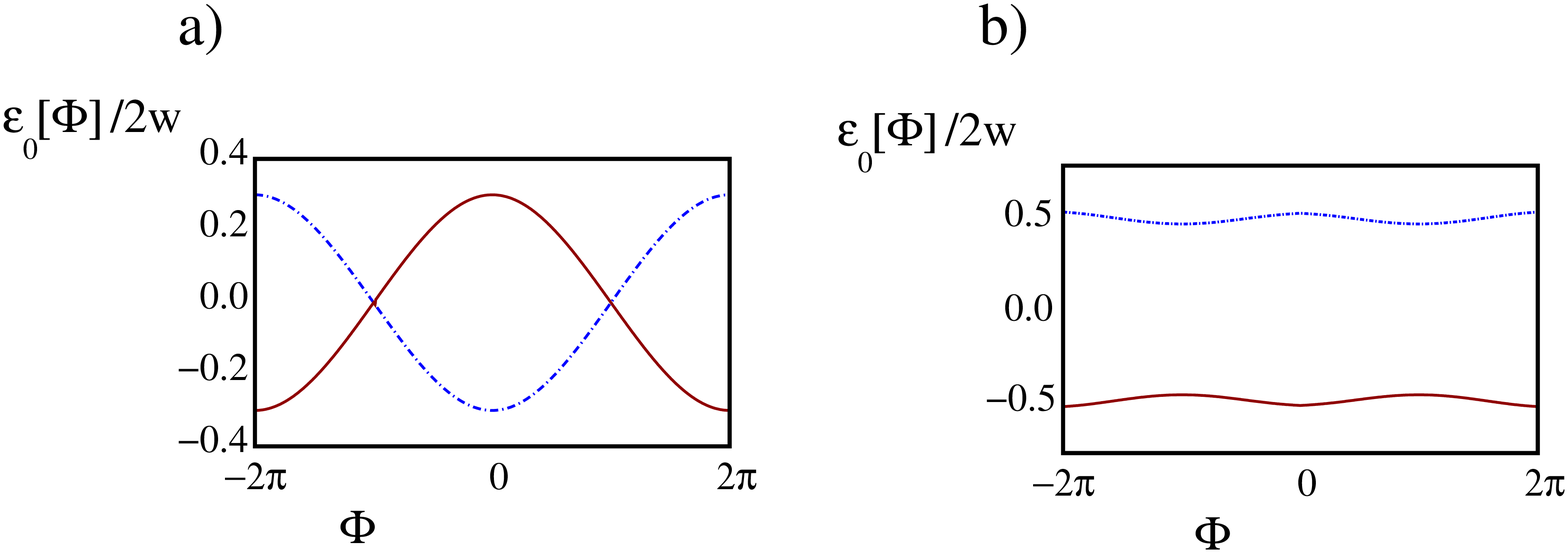}
\caption{Subgap energy levels $\pm \epsilon_0 [ \Phi ]$ as a function of the applied magnetic flux $\Phi$ 
for: \\ 
{\bf a)}: The p-wave superconducting ring with a weak link described by 
$H_K$ in Eq.(\ref{mh.1}) with $\Delta = w$ plus $H_\tau$ in Eq.(\ref{mh.2}) with $\frac{\mu}{2 w} = 0$, 
$\frac{\tau}{2 w} = 0.15$, and $\ell = 40$. At this value of the parameters, one recovers
a PMF LC at $\Phi_* = \pi$; \\
{\bf b)} The s-wave superconducting ring with a weak link described by 
$H_{s-wave}$ in Eq.(\ref{mh.5}) with $\Delta = w$ plus $H_{s-\tau}$ in Eq.(\ref{mh.6}) with 
 $\frac{ \mu}{2w} = 0$, $\frac{\tau}{2 w} = 0.15$, and $\ell = 40$. We again see a level modulation 
with $\Phi$, but now the levels emerge nearby the gap edge and keep far from the Fermi level at 
any value of $\Phi$.} \label{levels}
\end{figure}

\begin{figure}
\includegraphics*[width=1.\linewidth]{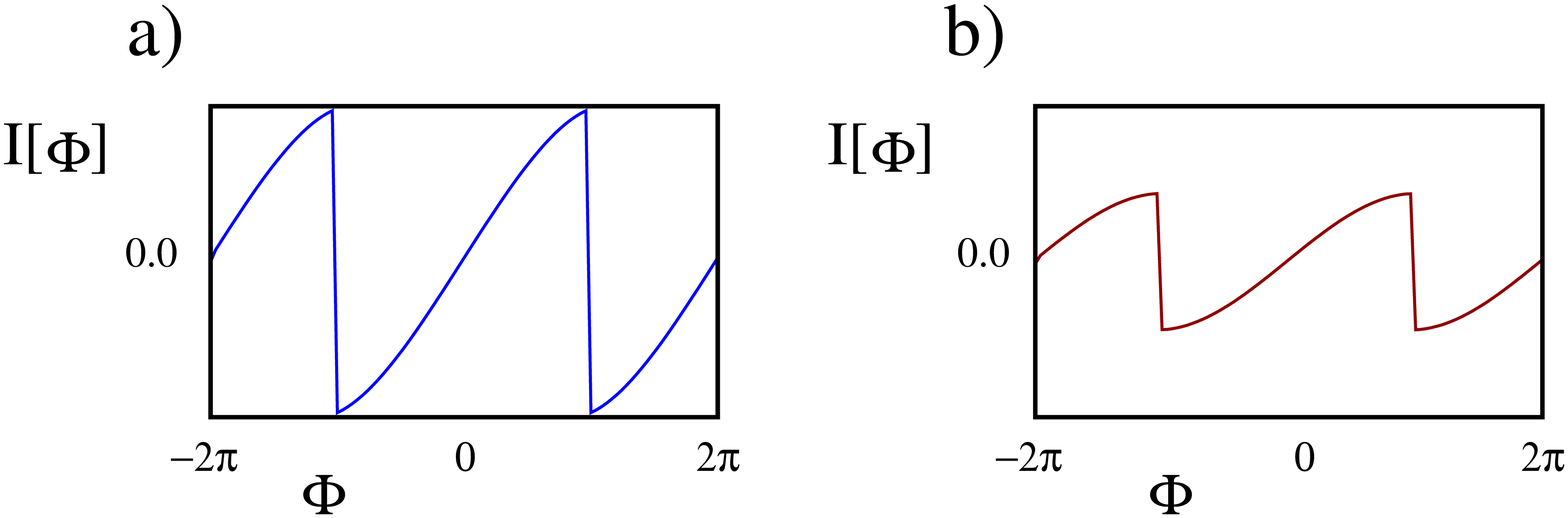}
\caption{Persistent current $I [ \Phi]$ (arbitrary units) {\it vs.}  $\Phi$ in a p-wave superconducting ring 
with $\ell = 40$, $\Delta = w$, a weak link of strength $\frac{\tau}{2 w}=0.15$, and 
with: \\
{\bf a)}: Chemical potential $\mu = 0$; \\  
{\bf b)} Chemical potential $\frac{\mu}{2 w}=0.75$.} \label{current_topo}
\end{figure}

Throughout our paper we assume that Eq.(\ref{condition}) is satisfied (differently 
from what is done in Ref.[\onlinecite{pientka}]), and that FP is not conserved. 
As a consequence of Eq.(\ref{condition}),  PMF energies $\pm \epsilon_0 [ \Phi ]$ {\it vs.} $\Phi$ 
show a sequence of LCs as displayed  in Fig.\ref{levels}a, where 
we plot the  PMF energies $\pm \epsilon_0 [ \Phi ]$ {\it vs.} $\Phi$ for 
a PSR with parameters chosen as outlined in the caption (by comparison, in 
Fig.\ref{levels}b we draw a similar diagram constructed for an 
s-wave superconducting ring described by the (spinful) Hamiltonian 
$H_s = H_{s-wave} + H_{s-\tau}$, with 
 
\beq
H_{s-wave} = - w \sum_\sigma \sum_{ j = 1}^{\ell - 1 } \{ c_{ j , \sigma}^\dagger 
c_{ j + 1 , \sigma } + c_{ j + 1 , \sigma}^\dagger c_{ j , \sigma} \} 
- \mu \sum_\sigma \sum_{ j = 1}^{\ell  }  c_{ j  , \sigma}^\dagger c_{ j , \sigma} 
+ \Delta \sum_{ j = 1}^\ell \{ c_{ j , \uparrow} c_{ j , \downarrow} + 
c_{ j , \downarrow}^\dagger c_{ j , \uparrow}^\dagger \}
\:\:\:\: , 
\label{mh.5}
\eneq
\noindent
and the weak link Hamiltonian given by 

\beq
H_{s-\tau} = - \tau \sum_\sigma \{ e^{ \frac{i}{2} \Phi} c_{ 1 , \sigma}^\dagger c_{ \ell , \sigma}
+  e^{ - \frac{i}{2} \Phi}  c_{ \ell , \sigma}^\dagger c_{ 1 , \sigma} \}
\:\:\:\: . )
\label{mh.6}
\eneq
\noindent
\noindent
To illustrate the consequences of the absence of FP conservation,  in  Fig.\ref{current_topo}, 
we plot  $ I [ \Phi ]$ {\it vs.} $\Phi$  in two PSRs in the PTP, 
with parameters set as in caption. In both cases,  we see  the 
discontinuity in $ I [ \Phi ]$ at $\Phi = \Phi_*$, with $\Phi_*$ depending on the system parameters 
as from Eq.(\ref{sg.47}).

\begin{figure}
\includegraphics*[width=1.\linewidth]{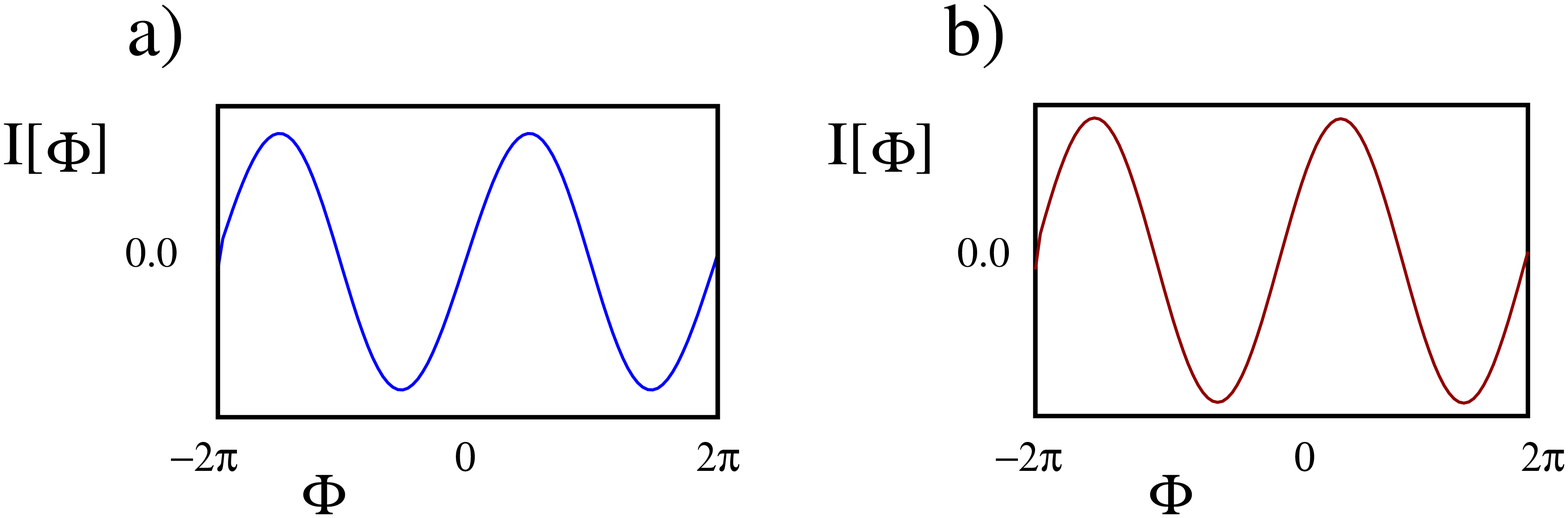}
\caption{\\ {\bf a)} Persistent current $I [ \Phi]$ (arbitrary units) {\it vs.}  $\Phi$ in an s-wave superconducting ring 
with $\ell = 40$, $\Delta = w$, a weak link of strength $\frac{\tau}{2 w}= 0.15$, and  
$\mu = 0$ ; \\  
{\bf b)} Persistent current $I [ \Phi]$ (arbitrary units) {\it vs.}  $\Phi$ in a p-wave superconducting ring 
with $\ell = 40$, $\Delta =  w$, a weak link of strength $\frac{\tau}{2 w}= 0.15$, and  
$\frac{\mu}{2 w}= 1.25$ . }  \label{nontopo}
\end{figure}

\begin{figure}
\includegraphics*[width=1.\linewidth]{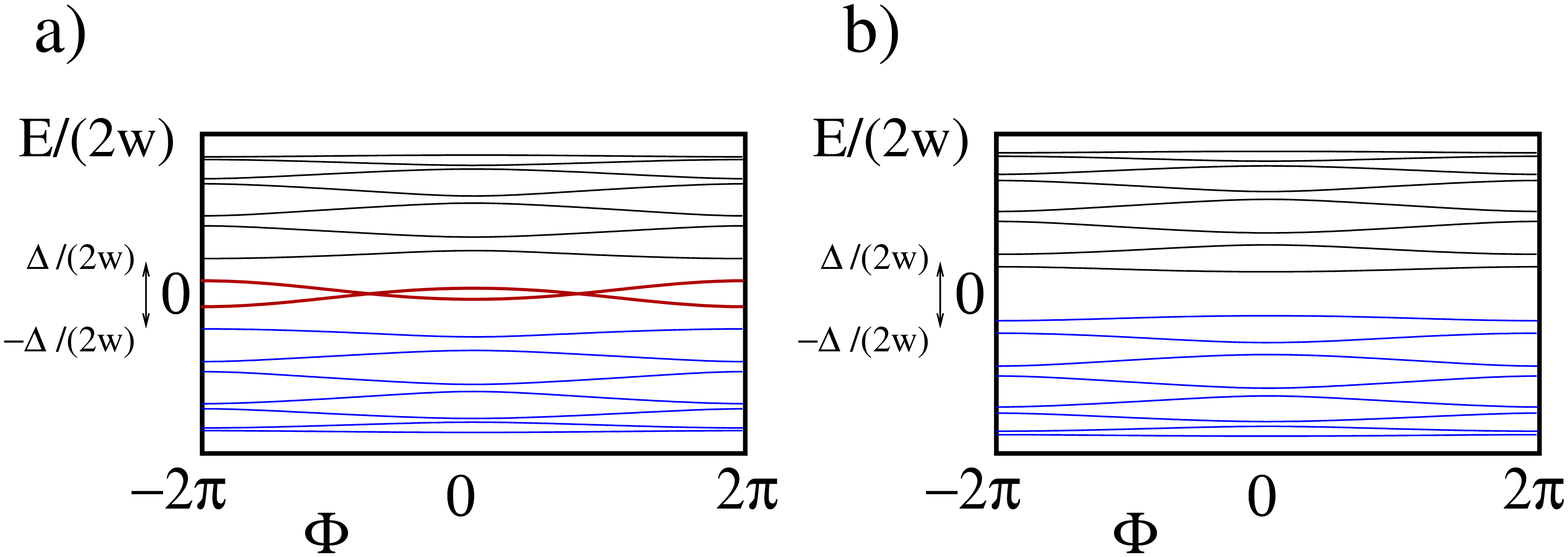}
\caption{\\ {\bf a)} Single-quasiparticle energy levels $ \epsilon_n [ \Phi ]$ (arbitrary units) {\it vs.}  $\Phi$  in a p-wave superconducting ring 
with $\ell = 8$, $\Delta = w$, a weak link of strength $\frac{\tau}{2 w}= 0.25$, and  
$\frac{\mu}{2 w} = 0.75$. The subgap PMFs appearing close to zero-energy are highlighted in red color; \\  
{\bf b)} Single-quasiparticle energy levels $ \epsilon_n [ \Phi ]$ (arbitrary units) {\it vs.}  $\Phi$  in a p-wave superconducting ring 
with $\ell = 8$, $\Delta = w$, a weak link of strength $\frac{\tau}{2 w} = 0.25$, and  
$\frac{\mu}{2 w}= 1.25$. Consistently with the discussion in the main text, no PMF-levels appear, in this 
case.}  \label{levels_phi}
\end{figure}

\noindent
For  comparison, in Fig.\ref{nontopo}, we plot $ I [ \Phi ]$ {\it vs.} 
$\Phi$, for a s-wave superconducting ring and  for a PSR not in the PTP:    in both cases 
$ I [ \Phi ]$ is a continuous function of $\Phi$, with no discontinuities within the whole 
interval $ [ - 2 \pi , 2 \pi ]$. 
\noindent
As a comment  about the discontinuity in $ I [ \Phi ]$, it is worth stressing that, 
while in the plots in Fig.\ref{current_topo}, we show the exact result obtained   
from Eq.(\ref{mh.8}) by summing over 
all the quasiparticle levels with $\epsilon_n < 0$, the discontinuity is 
clearly determined only by the   change in the slope of the  PMF energy at the LC.
In fact, this is a consequence of the fact that 
the levels with energy $ | \epsilon_n  | > \Delta_w$, together with 
their derivatives, are   
continuous functions of $\Phi$, so that they contribute $ I [ \Phi ]$ by a component
that is a smooth function of $\Phi$. To highlight this point, 
in Fig.\ref{levels_phi}a and in  Fig.\ref{levels_phi}b, we plot the energy levels {\it vs.} $\Phi$ 
respectively in a PSR in the PTP, and in a PSR not in the PTP. We see that, in the former plot, 
LCs at the Fermi level only concern the PMF. Instead, in the latter plot, there are no 
LCs at all, as expected. 
 
\noindent
The discontinuity of $I[ \Phi ]$ at $ \Phi = \Phi_*$ is a readily detectable feature that,
under minimal requirements on the system parameters, can be effectively used to mark the 
existence of a PMF LC by just  measuring $ I [ \Phi ]$ in a static experiment and looking at the specific dependence of 
the current on the applied flux. As we  are going to discuss in the following, 
this method it can be straightforwardly extended to study
to which extent a PMF LC survives in the presence of disorder and, ultimately, 
to map out the whole PTP.

\section{Low-energy subgap states in the presence of disorder}
\label{disorder_1}

In the previous section we show that, in the absence of disorder, in a pertinently 
engineered PSR  there always exists a flux $\Phi = \Phi_*$ at which, due to 
PMF LC, one recovers the zero-energy MMs   
$\gamma_1 , \gamma_2$ in  Eq.(\ref{ssg.4}) of Appendix \ref{closed_ring}. In this section, 
we discuss the robustness of the PMF LC against disorder in the PSR. It is by now
 established not only that the topological phase survives a moderate amount of 
disorder  in a disordered quantum wire in the presence of a strong spin-orbit coupling and 
at a large enough Zeeman effect \cite{brouwer_1,brouwer_2,pientka_1}, but also that a limited 
amount of disorder stabilizes the topological phase of an open, infinite Kitaev chain 
\cite{shuricht,wish}. Moreover, it was also stated that moderate disorder does not 
substantially affect the $4 \pi$-periodic component of $ I [ \Phi ]$ in a FP-conserving 
p-wave superconducting  ring with a weak link \cite{pientka}. Here, instead, we study to what 
extent the PMF LC in a FP non-conserving PSR is robust against disorder. 
Specifically,  we perform   a detailed analysis of the DOS and of the dependence of the energy 
levels on $\Phi$,  with particular emphasis onto the PMFs  in the presence of disorder.
As outlined in the following, we provide numerical evidence that, at a given disorder 
strength, either subgap PMF energy levels are still present, and they exhibit a LC at
a pertinent value of $\Phi$, or they are fully washed out by disorder. This 
leads us to the remarkable conclusion that, as long as PMFs are not washed out by 
disorder, looking at the  discontinuities  in $I [ \Phi ]$ is still
an effective way to probe a PMF LC, exactly as in the absence of 
disorder.

\begin{figure}
\includegraphics*[width=1.\linewidth]{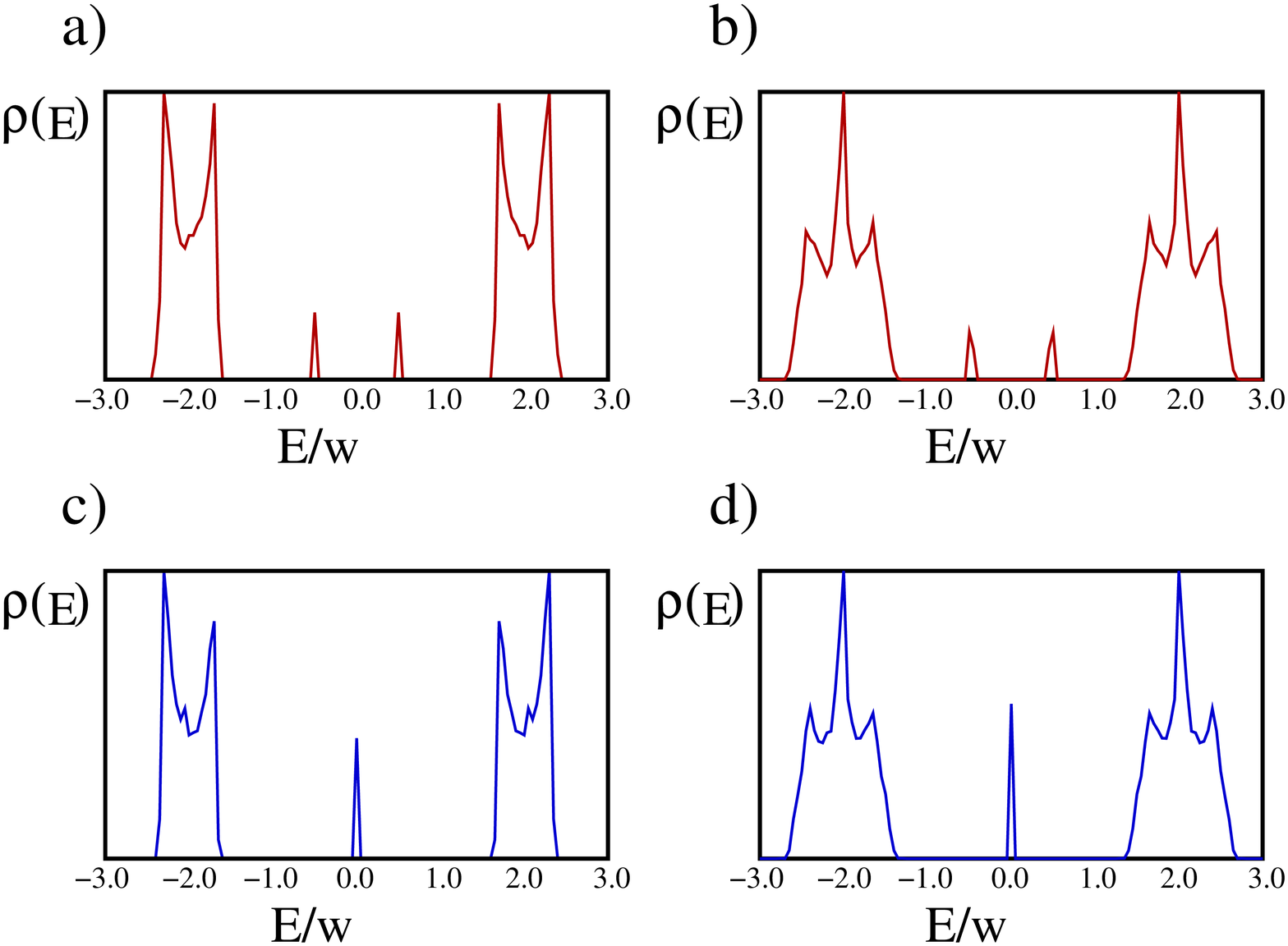}
\caption{\\ Single-quasiparticle DOS $\rho ( E )$ 
{\it vs.} $E$ for a ring with 
$\Delta = w$, $\frac{\mu}{2 w} = 0.15$ and $\frac{\tau}{2 w} = 0.25$, at a given 
value of the flux $\Phi$ and of $\sigma_V$. The 
remaining parameters used to generate the plots in the various panels 
have been set as outlined below: \\
{\bf a)} $\Phi$ = 0 , $\frac{\sigma_V}{w}= 0.05$ (a  tiny amount of disorder is added to the 
clean system  for the only purpose of regularizing the divergences at the poles 
of $\rho ( E )$ corresponding to the single-particle energy eigenvalues); \\  
{\bf b)} $\Phi = 0$ , $\frac{\sigma_V}{w} = 0.3$. At this value of  $\sigma_V$, one 
estimates  $l_0 \sim (2w  / \sigma_V)^2 > \ell$ \cite{pientka}: 
as expected  from the discussion given in  Ref.[\onlinecite{sau_1,sau_2}], the main effect of 
increased disorder are the emergence of a finite width for the PMFs and a slight 
renormalization of the effective superconducting gap; \\
{\bf c)} $\Phi = \pi$ , $\frac{\sigma_V}{w} = 0.05$. Same as in panel {\bf a)}, but now the two peaks
corresponding to the PMF levels have piled up into a taller peak, evidencing the 
PMF LC at the Fermi levels; \\
 {\bf d)} $\Phi = \pi$ , $\frac{\sigma_V}{w} = 0.3$. Same as in panel  {\bf b)}, but now the two peaks
corresponding to the PMF levels have piled up into a taller peak, evidencing the 
PMF LC at the Fermi levels.}  \label{dos}
\end{figure}

\begin{figure}
\includegraphics*[width=1.\linewidth]{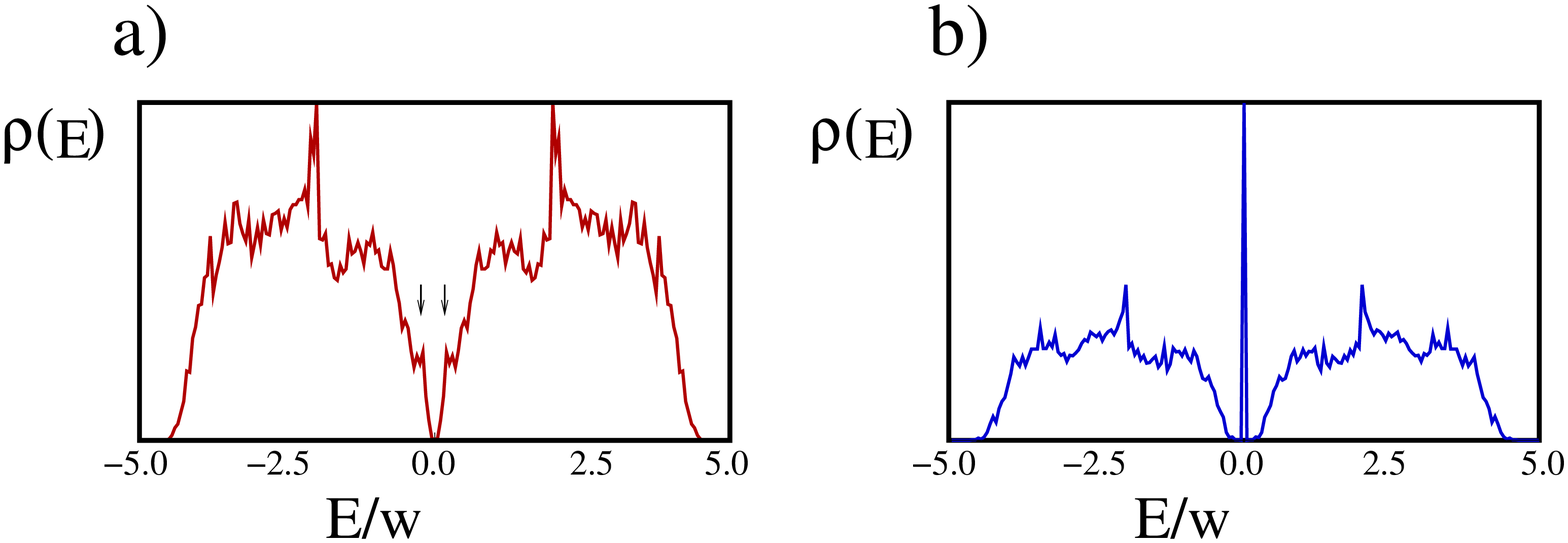}
\caption{\\ Single-quasiparticle DOS $\rho ( E )$ 
{\it vs.} $E$ for a ring with 
$\Delta = w$, $\frac{\mu}{2 w} = 0.15$ and $\frac{\tau}{2 w} = 0.25$, at a given 
value of the flux $\Phi$ and of $\sigma_V$. The 
remaining parameters used to generate the plots in the various panels 
have been set as outlined below : \\
{\bf a)} $\Phi$ = 0 , $\frac{\sigma_V}{w} = 1.5$ (the small black arrows highlight the 
peaks corresponding to the PMF energy levels); \\  
{\bf b)} $\Phi = \pi$ , $\frac{\sigma_V}{w} = 1.5$.}  \label{dos_15}
\end{figure}

To model the disorder, we modify the clean system Hamiltonian of 
Sec. \ref{model_hamiltonian} by adding a random component to
the on-site potential, so that, at a fixed disorder realization, 
the total Hamiltonian $H = H_K + H_\tau [ \Phi ] $ is modified  as 

\beq
H_K + H_\tau  [ \Phi ] \longrightarrow H_{ \{ V \} } [ \Phi ]  
\equiv H_K + H_\tau [ \Phi]  - \sum_{ j = 1}^\ell V_j c_j^\dagger c_j 
\:\:\:\: . 
\label{disl.1}
\eneq
\noindent
The  $\{ V_j \} $  are independent  random variables described by a probability 
distribution $ P [ \{ V_j \} ] = \prod_{ j = 1}^\ell p ( V_j )$, with 
$p ( V )$ being a probability distribution for $V$ with average 
$\bar{V} = \int \: d  V \: V p ( V ) = 0$, and with 
variance $\sigma_V^2 = \int \: d V \: V^2 p ( V ) $, which implies

\begin{eqnarray}
 && \overline{  V_j } = \int \: \prod_{ r = 1}^\ell d V_r \: P [ \{ V_r \} ] V_j = 0 \nonumber \\
 && \overline{ V_i V_j } = \int \: \prod_{ r = 1}^\ell d V_r \: P [ \{ V_r \} ] V_i V_j =
 \sigma_V^2 \delta_{ i , j } 
 \:\:\:\: , 
 \label{disl.2}
\end{eqnarray}
\noindent
with $\overline{ O [ \{ V_j \} ] }$ denoting the ensemble average of a generic functional 
of $\{ V_j \}$ with respect to the probability distribution $ P [ \{ V_j \} ]$. We  
use the uniform probability distribution given by   

\beq
p ( V ) \: = \:  \Biggl\{ \begin{array}{l}
\frac{1}{2 \sqrt{3} W} \;\; , \; {\rm for} \: - \sqrt{3}W \leq V \leq \sqrt{3} W \\
0 \:\: , \: {\rm otherwise}
                 \end{array}
\:\:\:\: , 
\label{disl.4}
\eneq
\noindent
which corresponds to setting $\sigma_V^2 = W^2$.  We now consider how disorder 
modifies the single-particle DOS of the ring. In general, for a finite-size system,
one expects that the effects of disorder strongly
depend on the ratio between the system size $\ell$ and the disorder-associated 
mean free path $l_0$.  In fact, a ''weak'' disorder, with 
$l_0 > \ell$, is expected to merely quantitatively affect the  DOS. In view of the analytical results 
obtained for an open chain  in Ref.[\onlinecite{sau_1}] within self-consistent Born approximation, later on 
numerically confirmed in Ref.[\onlinecite{sau_2}], we expect that 
a moderate disorder just slightly broadens the subgap peaks corresponding to 
the PMF energy levels and  
provides a possible slight renormalization of $\Phi_*$, without spoiling 
the  LC. To spell out the effects of increasing disorder strength, we     
numerically computed the exact single quasiparticle DOS $\rho ( E )$   at increasing  $\sigma_V$
for a PSR in the PTP  by using the Hamiltonian in Eq.(\ref{disl.1}) at fixed $\{ V_j \}$.
Therefore,  we  ensemble-averaged  the result over 300 realization of the disorder. 
We show the results in Fig.\ref{dos}, where one clearly sees the expected broadening of 
the subgap peaks, as $\sigma_V$ increases.  
\noindent
At  strong disorder ($l_0 \ll \ell$), the energy levels of the system 
become distributed according symmetry class-D  random
matrix distribution  \cite{hedge,Beenakker_review,alt_zirn}. The 
corresponding  level statistics is given by 

\beq
{\cal P} [ \{ \epsilon_j \} ] \prod_j \: d \epsilon_j \: \propto  
\prod_{ i < j } | \epsilon_i^2 - \epsilon_j^2 |^\beta \prod_k [ | \epsilon_k |^\alpha \:
e^{ - \frac{\epsilon_k^2}{\sigma_V^2}}
\: d \epsilon_k ] 
\:\:\:\: , 
\label{disl.3}
\eneq
\noindent
with $\beta = 2$ and  $\alpha = 0$ and the product taken over positive-energy levels only. 
From Eq.(\ref{disl.3}) it  is possible 
to extract the single-particle DOS as $\epsilon \to 0$, given by \cite{alt_zirn}  
$\rho ( \epsilon ) \propto | \epsilon |^\alpha$.
Thus, for symmetry class-D one expects a low-energy uniform DOS, with no evidence 
of low-lying PMFs. An enlightening discussion of how the disorder washes out PMFs is provided in 
Ref.[\onlinecite{motrunich}]: assuming that at zero disorder the system lies within its topological 
phase, one finds that, for weak disorder, fluctuations in the random potential may open ''nontopological
islands'' within the topological background, of typical size $\ell_{\rm NT} \ll \ell$. At each interface 
between topological and nontopological regions, MMs emerge, which suddenly hybridize into ''high-energy 
Dirac modes'', at typical energy scales $\epsilon_{NT} 
\sim e^{ - \frac{l_0}{\xi_0}}$, lying below the gap but still
higher than the typical energy associated to the ''true'' PMFs. 
On one hand, this implies the 
proliferation of subgap, disorder-induced states. On the other hand, low-energy PMFs are still
protected by the very fact that their energies are quite lower than 
the energies associated to disorder-induced
states, and such is the corresponding LC at $\Phi_*$. Summarizing, as 
$\sigma_V$ increases, one legitimately 
expects that disorder-induced energy levels fill in the subgap region but 
also that level repulsion between
states with different energy, encoded in ${\cal P} [ \{ \epsilon_j \} ]$ in Eq.(\ref{disl.3}), 
acts to ''protect'' the low-lying PMFs. This is clearly evidenced by the DOS plots we provide in 
Fig.\ref{dos_15}, where we show $ \rho ( E )$ for a PSR with the same parameters we used to generate Fig.\ref{dos},
but at $\sigma_V$ such that $l_0 / \ell \sim 0.1$. We see that, while 
disorder-induced levels largely fill in the energy 
gap up to the PMF levels, the regions between the two peaks in  Fig.\ref{dos_15}a 
are still empty: a signal that the PSR is still within the PTP. Similarly, 
the persistence of this central peak in Fig.\ref{dos_15}b due to PMF pile up
evidences that the PMF LC at $\Phi = \Phi_*$ is not spoiled by disorder, 
even at $\sigma_V$ as large as $1.5 w$, consistently with the general expectation 
about class-D symmetry models  \cite{bouquet,mehta,been_2014}.

\begin{figure}
\includegraphics*[width=1.\linewidth]{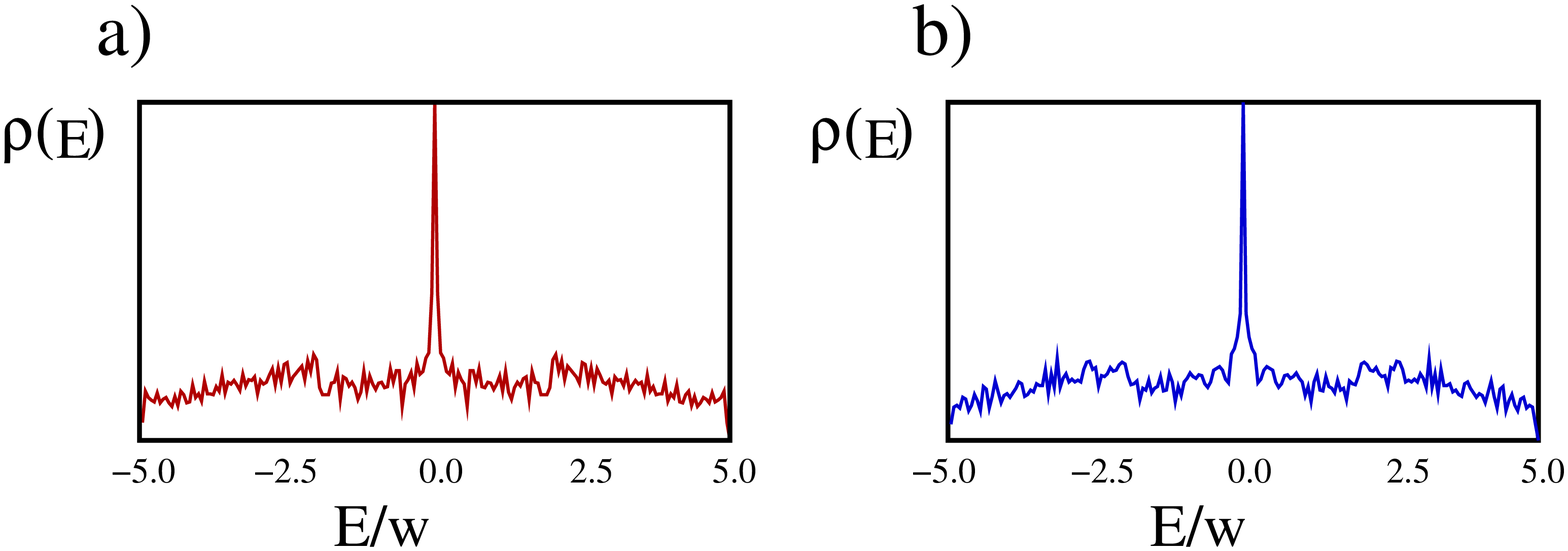}
\caption{\\ Single-quasiparticle DOS $\rho ( E )$ 
{\it vs.} $E$ for a ring with 
$\Delta = w$, $\frac{\mu}{2 w} = 0.15$ and $\frac{\tau}{2 w} = 0.25$, at a given 
value of the flux $\Phi$ and of $\sigma_V$. The 
remaining parameters used to generate the plots in the various panels 
have been set as outlined below: \\
{\bf a)} $\Phi$ = 0 , $\frac{\sigma_V}{w} = 4.0$ ; \\  
{\bf b)} $\Phi = \pi$ , $\frac{\sigma_V}{w} = 4.0$.}  \label{strong_disorder}
\end{figure}

\begin{figure}
\includegraphics*[width=1.\linewidth]{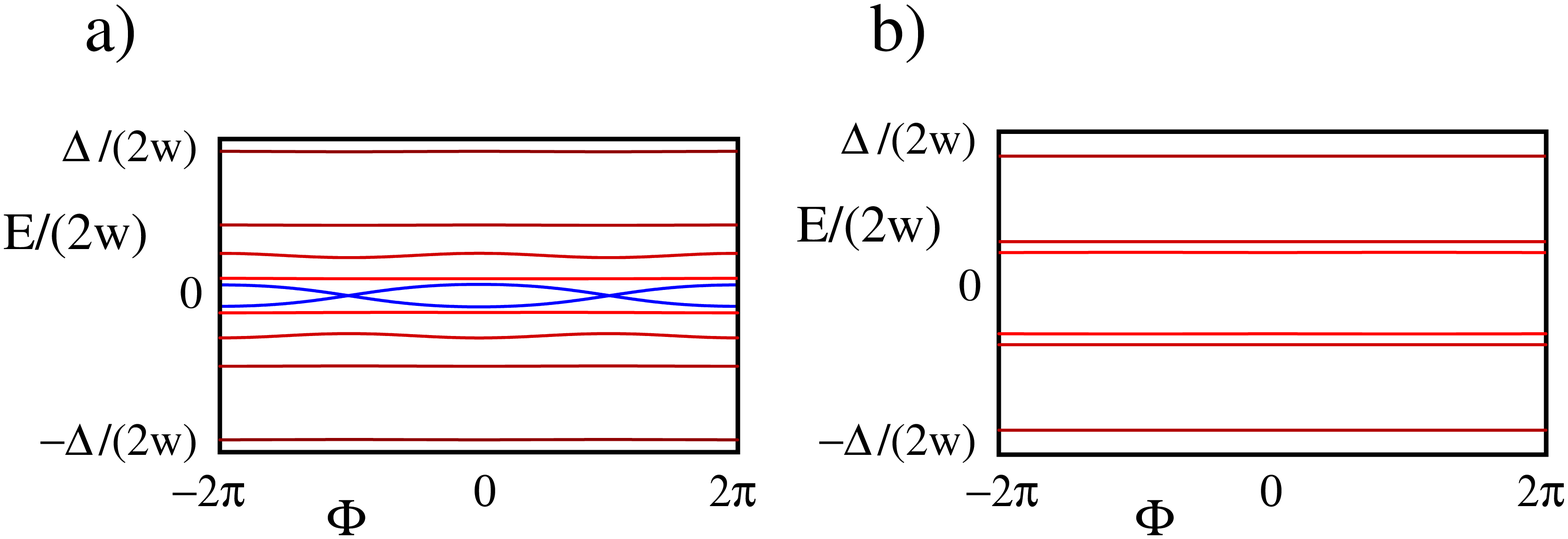}
\caption{\\ {\bf a)} Sub-gap energy levels   $ \epsilon_n [ \Phi ]$  computed for one 
realization of the disorder potential in a p-wave superconducting ring 
with $\ell = 40$, $\Delta = w$, a weak link of strength $\frac{\tau}{2 w} = 0.25 $,   
$\frac{\mu}{2 w} = 0.75$, and   $\frac{\sigma_V}{w} = 2.2$. The subgap PMFs appearing close 
to zero-energy are highlighted in blue color; \\  
{\bf b)} Sub-gap energy levels   $ \epsilon_n [ \Phi ]$  computed for one 
realization of the disorder potential in a p-wave superconducting ring 
with $\ell = 40$, $\Delta = w$, a weak link of strength $\frac{\tau}{2 w} = 0.25 $,   
$\frac{\mu}{2 w} = 0.75 $, and   $\frac{\sigma_V}{w}  = 4.2$. No subgap PMFs appear in this 
case. }  \label{disorder_a}
\end{figure}

\noindent
In Fig.\ref{strong_disorder} we plot $ \rho ( E )$ for a PSR with the same parameters as 
the ones considered above, but with $\sigma_V / w = 4.0$. On comparing  Fig.\ref{strong_disorder}a
to  Fig.\ref{strong_disorder}b, we see no appreciable differences, that is, we  find 
no detectable differences between the plots at $\Phi = 0$ and at $\Phi = \pi$. 
In fact, the insensitivity of the DOS to the applied flux can be regarded as a specific manifestation of the insensitivity 
to the boundary conditions (such as the one imposed by the applied flux 
$\Phi$ on the single-quasiparticle
wavefunction) of the energy levels corresponding to localized states 
\cite{thouless_1,thouless_2}. This observation leads us to the conclude 
that all the states filling in the gap at strong disorder 
are disorder-induced states, while strongly $\Phi$-sensitive PMFs have been completely 
washed out. To further ground our conclusion, in   Fig.\ref{disorder_a}a, we plot 
the first quasiparticle energy levels of the PSR $ \epsilon_n [ \Phi ]$ {\it vs.} $\Phi$  
at  $\sigma_V /  w  = 2.2$ for a fixed disorder configuration. We clearly see a set of 
disorder-induced states (drawn in red in the figure) which exhibit a 
weak dependence on $\Phi$ and  are situated symmetrically with respect to  
0 energy, consistently with the survival of particle-hole symmetry against
disorder.  The states closest to the  Fermi level (depicted in blue in 
the figure) are, instead, to be clearly identified 
with PMFs. They take a strong dependence on $\Phi$ and cross with each other at 
pertinent values of $\Phi$. In Fig.\ref{disorder_a}b, we draw 
a similar plot, but realized for   $\sigma_V / w  = 4.2$. The much larger amount of 
disorder has  determined the full disappearance of PMFs \cite{shuricht}: there are 
no blue states, but only red impurity states, basically independent of $\Phi$ (a clear
signal of strong localization of these states).

\noindent

\noindent
The peak centered at $E =0$ 
in the plots of Fig.\ref{strong_disorder} corresponds to 
Griffith's singular behavior in the DOS cutoff at the finite 
level spacing $\delta_0 \sim 2 \pi w / \ell$ \cite{been_2014}. In fact, 
on increasing $\sigma_V$, the disorder washes out the PMFs via the Griffiths effect 
taking place in the finite wire \cite{motrunich,sau_2}. When the nontopological regions 
start to proliferate, the increasing probability of hybridization between PMFs and 
zero-modes located at the interfaces between topological and nontopological regions 
eventually washes out the PMFs themselves, together with the degenerate point at $\Phi = \Phi_*$, 
driving the system outside of the PTP \cite{motrunich,sau_2}.

\section{Putative topological phase boundaries and discontinuities in the persistent current}
\label{PTP_boundaries}

\begin{figure}
\includegraphics*[width=1.\linewidth]{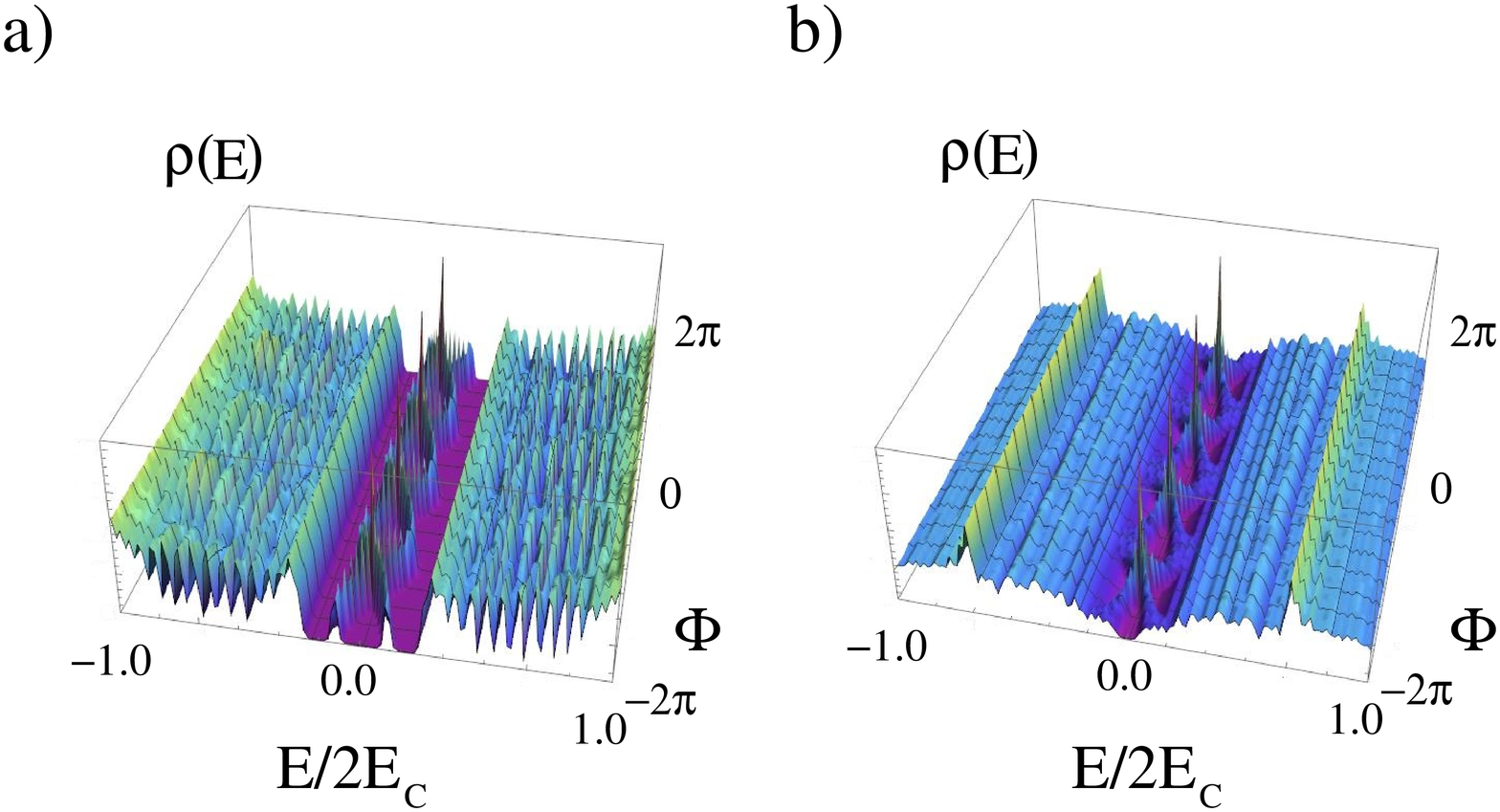}
\caption{\\ {\bf a)} DOS $\rho ( E )$ for a disordered ring 
described by the Hamiltonian in Eq.(\ref{disl.1}) with $\Delta = w$, $\frac{\mu}{2 w} = 
1.2$, $\frac{\tau}{2 w} = 0.25$,  $\ell = 40$, obtained by ensemble-averaging over 100 realization of the disorder with 
$\frac{\sigma_V}{w}  = 0.2$. The subgap PMFs close to the Fermi level are 
clearly displayed; \\
 {\bf b)} DOS $\rho ( E )$ for a disordered ring 
described by the Hamiltonian in Eq.(\ref{disl.1}) with $\Delta = w$, $\frac{\mu}{2 w} = 
1.2$, $\frac{\tau}{2 w} = 0.25$, $\ell = 40$,  obtained by ensemble-averaging over 100 realization of the disorder with 
$\frac{\sigma_V}{w}  = 0.8$. The subgap PMFs are less resolved, but there are clear peaks 
at the level crossings, due to the pile-up of the DOS of the 
two PMFs. These peaks survive the transfer of spectral weight from the  PMFs to disorder-induced Griffiths states,
which evidences that the PSR still lies within the PTP.    }  \label{three_dimensional}
\end{figure}

To discuss the correspondence between PMF LCs and discontinuities in 
$I  [ \Phi ]$ in a disordered PSR, we look at the  ensemble averaged DOS at fixed 
$\sigma_V$ as a function of $\Phi$, $\rho ( E ) $.  To  numerically
construct $\rho ( E ) $, we collect the eigenvalues  generated via 
an exact Hamiltonian diagonalization procedure into bins defined in the $E$-$\Phi$ plane and 
eventually average over the disorder with $p ( V )$ given in Eq.(\ref{disl.4}). 
We thus generate  three-dimensional plots of  $\rho ( E ) $ 
as a function of both $E$ and 
of $\Phi$ for $E$ ranging throughout the interval $[-E_C , E_C]$, with the half-bandwidth 
$E_C =  2 w + \mu $ and $\Phi \in [ - 2 \pi , 2 \pi ]$. These are reported in 
Fig.\ref{three_dimensional} for a PSR in the PTP at limited disordered.
In Fig.\ref{three_dimensional},  despite the presence of disorder, we clearly see the subgap PMFs, which are 
characterized by their strong dependence on $\Phi$ (to be contrasted with the
observations that all the other levels displayed in the figure are basically independent
of $\Phi$) and, more importantly, that there are evident LCs, evidenced by the sharp peaks - 
a consequence of the two PMFs density  pile-up at those points-. The LCs are not  washed out by disorder, which 
implies that, in a sense that we are going to clarify in the following, 
there is still a sort of discontinuous behavior of  $  I [ \Phi ]$ at  $\Phi = \Phi_*$.

\begin{figure}
\includegraphics*[width=.4\linewidth]{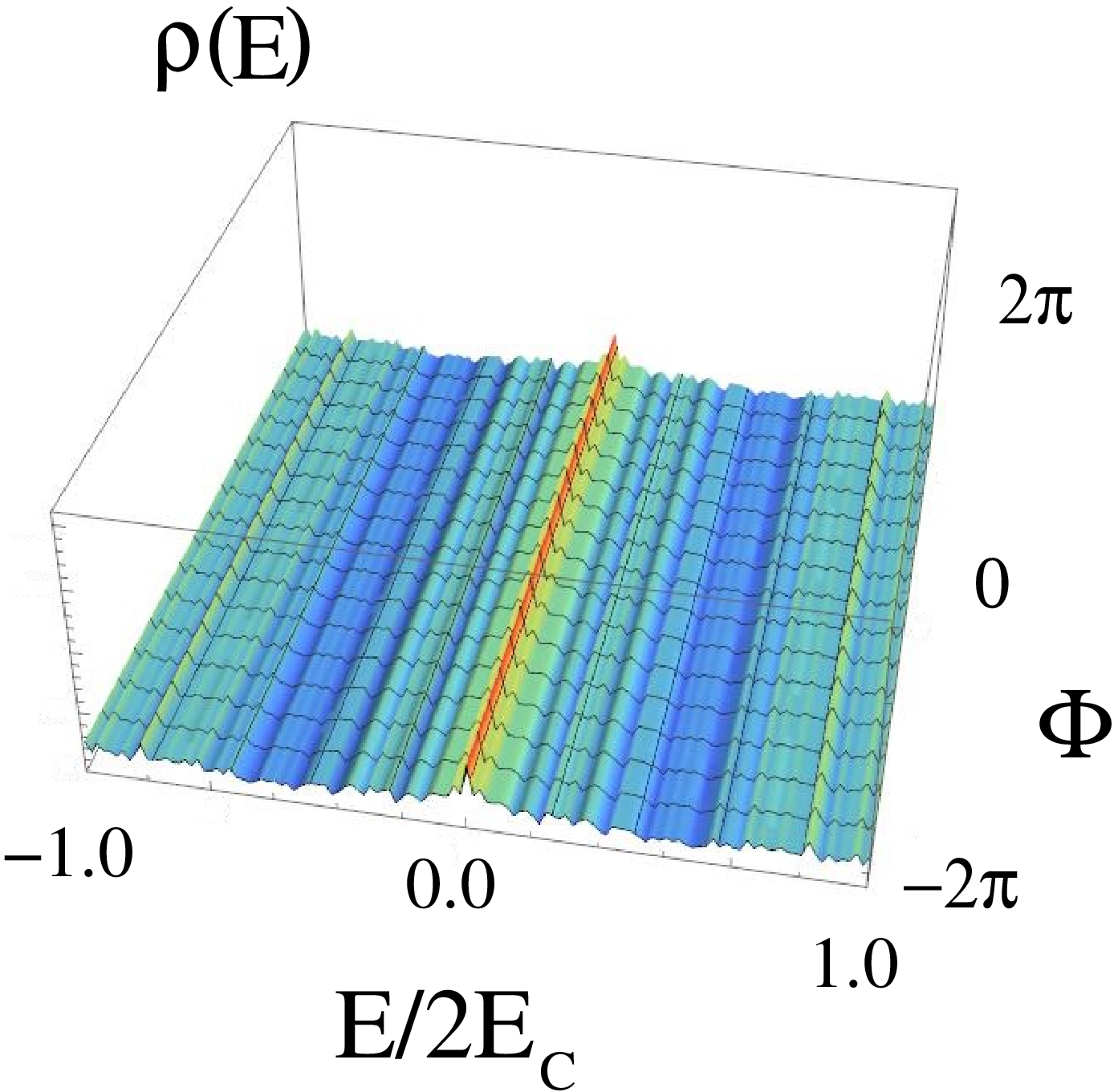}
\caption{\\ DOS $\rho ( E )$ for a disordered ring 
described by the Hamiltonian in Eq.(\ref{disl.1}) with $\Delta = w$, $\frac{\mu}{2 w} = 
1.2$, $\frac{\tau}{2 w} = 0.25$, $\ell = 40$, obtained by ensemble-averaging over 100 realization of the disorder with 
$\frac{\sigma_V}{w}  = 6.0$. The   spectral density is 
 broadened throughout the interval $[ - E_C , E_C ]$, there is a negligible 
dependence on $\Phi$, and the subgap PMFs have disappeared. }  \label{large_disorder}
\end{figure}

\noindent
By contrast, in Fig.\ref{large_disorder}, we  plot  $\rho ( E )$ with the same parameters as 
we chose in Fig.\ref{three_dimensional}, but with $\sigma_V / w  = 6.0$. We see that the spectral weight is 
largely broadened throughout the interval $[ - E_C , E_C ]$, there is no dependence of  the energies on 
 $\Phi$ and, at the same time, the subgap PMFs have disappeared. This  confirms the conclusion we 
 reached in Sec. \ref{disorder_1}, that  a strong enough disorder is actually effective in  washing  out the 
 subgap PMFs.  

\begin{figure}
\includegraphics*[width=.6\linewidth]{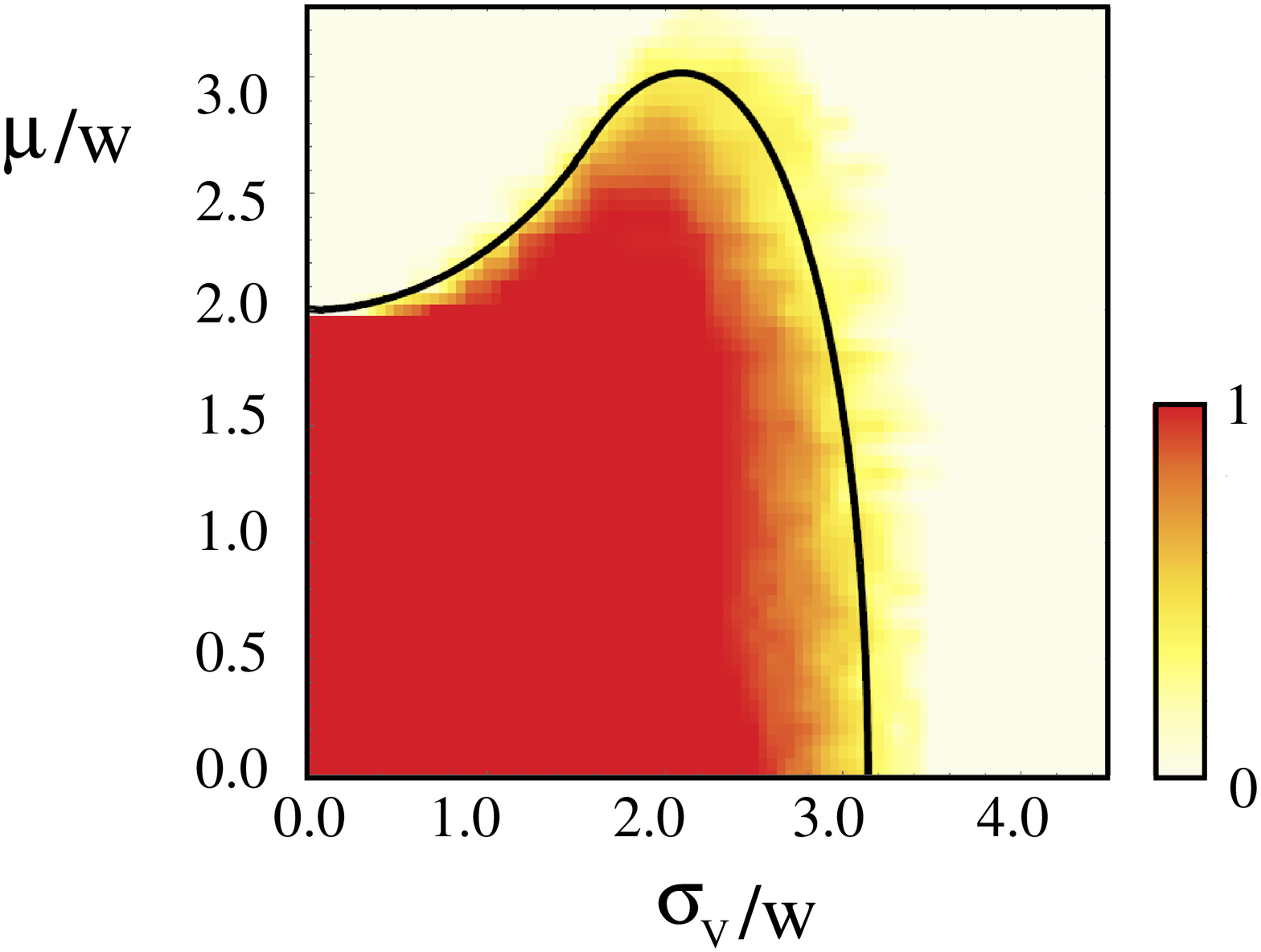}
\caption{ Color-scale plot of $f [ \sigma_V , \mu ]$ in the $\sigma_V - \mu$ plane. The color 
code is summarized at the right-hand side of the plot. The shaded 
region, where the color varies from red to white , defines the transition region at which the 
PTP disappears and the PMFs are washed out of the spectrum. The red line is a sketch of the variation of 
the center of mass of the transition region as a function of $\mu$: in the $\ell \to \infty$ limit 
and after averaging over a large number of realizations of disorder, it is expected to coincide with 
the solid black line of Fig.1 of Ref.[\onlinecite{shuricht}].}  \label{phase_diagram}
\end{figure}

\noindent
To map out the PTP, we  generalize to a disordered PSR the correspondence between PMF LCs and discontinuities in
$I  [ \Phi ]$. We first divide the region of interest in the $\sigma_V - \mu$ planes into square bins and 
define at the center of each bin a function $F [ \sigma_V , \mu ]$ which, at the start, is everywhere equal to
0. Then, at each bin $ ( \sigma_{V , i}, \mu_j )$, we  extract a realization $ \{ V_j \}$ of 
the disorder with probability having $\sigma_V = \sigma_{ V , i }$ and exactly diagonalize 
the Hamiltonian  $H_{ \{ V \}  ; \mu_j } [ \Phi ]$ obtained from Eq.(\ref{disl.1}) by setting  $\mu = \mu_j$.
Then, we use the result   to compute the corresponding persistent current, $I_{ \{ V \} } [ \Phi ]$. Therefore,
we check   whether $ I_{ \{ V \} } [ \Phi ]$ exhibits a discontinuity at some value $\Phi = \Phi_*$, or not.
If yes, we increment  $F [ \sigma_{V , i} , \mu_j ]$ by 1, otherwise, we leave it unchanged. Specifically, 
to identity $\Phi_*$ we discretize the interval of values of $\Phi$ and, starting from 
$\Phi = 0$, at each step we increment $\Phi $ by $\delta \Phi$, with $0 <  \delta \Phi \ll 2 \pi$. Setting $ \Phi_r = r \delta \Phi$, we 
look at the sign of the product $I [ \Phi_r ] I [ \Phi_{ r + 1 } ]$. As soon as $I [ \Phi_r ] I [ \Phi_{ r + 1 } ] < 0$,
we temptatively identify $\Phi_*$ with the middle point of the interval $ [ \Phi_r , \Phi_{ r + 1} ]$. We then 
numerically estimate the slope of $I [ \Phi ]$ at $\Phi_*$ as $ ( I [ \Phi_{ r + 1} ] - I [ \Phi_r ] ) / \delta \Phi$
and at the left-hand side of  $\Phi_*$ as  (  $ I [ \Phi_{ r } ] - I [ \Phi_{r-1} ] ) / \delta \Phi$, 
concluding that $\Phi_*$ corresponds to a discontinuity point if the two slopes differ from each other 
by a factor $\geq 1.5$. After summing, at each bin, over 
 ${\cal N} = 300$ random realizations of the disorder, we define  $f [ \sigma_{V , i} , \mu_j ]
 = F [ \sigma_{V , i} , \mu_j ] / {\cal N}$,  so that $0 \leq f [ \sigma_{V , i} , \mu_j ] \leq 1$, $\forall i , j $. 
As a final result, we draw the diagram  in Fig.\ref{phase_diagram}, where we show a color-scale plot of 
$ f [ \sigma_V , \mu ]$ computed  for a ring with $\ell = 60$, $w = \Delta$, and $\tau / (2w )= 0.25$. 
In detail, we constructed the plot 
by increasing both $\sigma_V / w$ and $\mu / w$ by step of 0.05 and by accordingly defining the  bins in the 
$\sigma_V - \mu$-plane. The region marked in full red corresponds to $f = 1$, that is, to a current which is singular at 
some $\Phi_*$ for any realization of the disorder. Conversely, the white portion of the graph corresponds to 
$f = 0$, that is, to a current that is a continuous function of $\Phi$ for any realization of the disorder
(see  Fig.\ref{phase_diagram} for  a graphic summary of the color code).
Clearly, points in the red region are characterized by PMFs that undergo a LC at $\Phi = \Phi_*$, that is, we 
may identify the red region with the PTP in the presence of disorder. By converse, points in the white region 
are characterized by the absence of PMFs. 
The shaded  area, where the color varies from red to white, defines the transition region at which the 
PTP disappears and the PMFs are washed out of the spectrum. By analogy, one would expect a sharp 
transition line, such as the one separating the topological from the nontopological phases of 
the infinite Kitaev chain, drawn in Refs.[\onlinecite{shuricht,wish}] by using  transfer matrix 
method. Nevertheless, we obtain a broad transition region, rather than a sharp phase boundary, because,  
for each disorder realization we exactly diagonalize a well-defined Hamiltonian, which either 
presents PMFs with a finite-$\Phi$ LC, or not. Near the phase transition, when averaging over ${\cal N}$ 
different realization of the disorder, there can  be a nonzero probability that 
some realizations wash out PMFs at values of the system parameters where PMFs are 
present in the large majority of cases or, conversely, that PMFs appear at points
in parameter space where they are absent in the large majority of cases. To be more 
precise, when the system lies within the PTP, strong fluctuations in the mean value of the impurity potential on a single 
realization of disorder may drive it outside of the PTP, and vice versa. 
As a result, close to the point of disappearance of the PTP, we expect 
that, on ensemble averaging over disorder, the percentage of single 
disorder realizations respectively leading to a discontinuous, or to 
a continuous, current will be both different from zero (incidentally, we 
note that this is quite a common feature of finite system undergoing a 
Griffith phase transition \cite{sau_2}). Instead, far from the transition
there is no ambiguity in that either $ I [ \Phi ]$ is discontinuous, or 
continuous for any realization of disorder, just as we can see from the plot 
of $ f [ \sigma_V , \mu ]$ in Fig.\ref{phase_diagram}. Eventually, to map out the
(broad) phase boundary of the PTP in the disorder strength-chemical potential 
plane, we  start from within the PTP at zero disorder strength. Then,  we 
move along the horizontal axis at fixed $\mu$ by probing the existence of the discontinuity in 
$ I [ \Phi ]$ at increasing $\sigma_V$: for $0 \leq \mu / w  < 2$ and for $\sigma_V = 0$, we typically obtain 
$ f [ 0 , \mu ] = 1$. Consistently with the above discussion, at some $\mu$-dependent 
''lower critical'' value of $\sigma_V$, $\sigma_{\mu ; l}$,  
we start to obtain $ f [ \sigma_{V ; l}  , \mu ] < 1$. This signals the start of the transition 
region when going across which the PTP disappears. On further increasing $\sigma_V$, one 
typically reaches an ''upper critical'' value, $\sigma_{ \mu ; u}$, such that 
$f [ \sigma_V ; \mu ] = 0 $ for $\sigma_V > \sigma_{ \mu ; u}$. Therefore, $\sigma_{ \mu ; l}$ and 
$\sigma_{ \mu ; r }$ determine the transition region at a given $\mu$ as the set of 
points $ (\sigma_V , \mu )$ such that $\sigma_{ \mu ; l } < \sigma_V < \sigma_{ \mu ; u}$. 
Repeating the procedure along constant-$\mu$ lines, we mapped out the full color scale 
plot of $ f [ \sigma_V , \mu ]$ in Fig.\ref{phase_diagram}.

Besides the broadening related to the Griffith mechanism, 
 Fig.\ref{phase_diagram} shows a remarkable analogy with
the phase diagram for a long  Kitaev chain with open boundary conditions reported 
in Fig.1 of Ref.[\onlinecite{shuricht}].  In particular, the two diagrams share 
the remarkable feature of a reentrant topological phase at not-too-large values of 
$\sigma_V$, that is, a small amount of disorder appears to favor, rather than suppressing,
the topological phase.   In our specific finite-size 
ring, we interpret the reentrant phase as an effect of the disorder-induced renormalization
of the chemical potential which, at nonzero $\sigma_V$, pushes the phase transition to 
values of $\mu$ higher than the zero-disorder critical value $\mu_c = 2 w $, or lower than 
$\mu_c = - 2 w $ (a detailed discussion of this effect, both using the reduced effective 
low-energy Hamiltonian for the ring and in general, as a consequence of turning on 
a weak disorder, is provided in Refs.[\onlinecite{pientka,pientka_1}]). As discussed above, the 
finite width of the transition region has to be regarded as a consequence of the Griffiths mechanism 
in a finite system \cite{sau_2}: in the $\ell \to \infty$ limit 
and after averaging over a large number of realizations of disorder, the transition region 
is nevertheless expected to shrink to a sharp phase boundary, coinciding  with 
the solid black line of Fig.1 of Ref.[\onlinecite{shuricht}].

\noindent

Fig.\ref{phase_diagram} summarizes the key results of this paper: on one hand, it can be 
regarded as a theoretical derivation of the region, in the $\sigma_V - \mu$ plane, in which 
it is possible to make two zero-energy MMs emerge at the quantum ring by pertinently 
acting on the applied flux. On the other hand, the way we derived it suggests a practical tool 
to map it out in an actual experiment:  to check whether, at a given values of the system parameters, 
zero-energy MMs are recovered at a disordered PSR,  it is enough
to probe the dependence of $I [ \Phi ]$ on $\Phi$ and to check whether, at some flux $\Phi = \Phi_*$
(with $\Phi_*$ typically being $\sim \pi$ for a not-too-short chain), $ I [ \Phi ]$ shows a discontinuous
behavior, with a finite jump when going across $\Phi_*$. A key point of our proposed  technique 
is that   $ I [ \Phi ]$ can in principle be probed by means of a noninvasive  magnetometer, without need 
for contacting the system as one has to do in a transport experiment attempting to probe MMs \cite{exp_1}; 
this potentially should avoid problems related to the introduction of   sources of noise in the system.  
  Moreover, we recall that the techniques so far proposed to 
map out the phase diagram of a disordered Kitaev-like chain, mostly rely on looking at the eigenvalues 
of the transfer matrix of the whole chain \cite{shuricht,wish}: an approach rigorous and effective from 
the mathematical point of view, but lacking the possibility of a direct experimental 
implementation.  At variance, as stated above, in order for our approach to work in 
practice, one simply needs to perform a noninvasive magnetic probe, thus providing a 
direct means to measure the emergence of MMs in the ring at an appropriate value of 
the applied flux. 

From the physical point of view, our numerical findings prove that  zero energy MMs emerging 
in a quantum ring at $\Phi = \Phi_*$ are quite robust against disorder. It would be 
extremely interesting to perform a rigorous  analytical investigation of the 
features of the PMFs and of the stability of the MMs at $\Phi = \Phi_*$: this 
topic lies nevertheless beyond the scope of this work and we plan to investigate 
it in a forthcoming publication.

\section{Conclusions and further perspectives}
\label{concl}

In this paper, we discuss the emergence and a way for probing  MMs at a disordered finite-length p-wave 
one-dimensional superconducting ring, pierced by a magnetic flux $\Phi$, in the absence of FP conservation. 
We prove  that, at a moderate amount of disorder,  it is still possible 
to tune $\Phi$ at a value $\Phi_*$ at which the subgap modes appear exactly at 
zero energy, due to the level crossing between the subgap energy levels. At $\Phi = \Phi_*$, the 
MMs are recovered as pertinent linear combinations of the subgap mode operators. 
To probe the level crossing, we propose to look for discontinuities in  the persistent current  $ I [ \Phi ]$
at  $\Phi = \Phi_* \sim \pi$. On pertinently employing our technique, we map out the whole region, in 
the disorder strength-chemical potential plane, characterized by zero-energy subgap LCs. At this point, 
we expect it to be possible to realize in a controlled and tunable way MMs which, in principle, provide 
a crucial and effective resource for designing efficient quantum computation protocols. 

In the presence of disorder, our results mostly rely on a numerical analysis: while it would be 
extremely interesting to perform a rigorous  analytical investigation of the 
features of the PMFs and of the stability of the MMs at $\Phi = \Phi_*$, this 
is outside the  scope of this work and we plan to investigate 
it in a forthcoming publication.

 \vspace{0.5cm}

  G. Campagnano acknowledges    financial support from MIUR-FIRB 2012 project ”HybridNanoDev” (Grant No. RBFR1236VV). 
  We acknowledge insightful discussions with A. De Martino, L. Lepori, P. Lucignano, A. Romito, and A. Tagliacozzo. 

\appendix
\section{Excitation spectrum of the open Kitaev chain with 
$w = \Delta$ }
\label{open_chain}

In this appendix we review the derivation of the  excitation spectrum of 
the  Kitaev Hamiltonian in Eq.(\ref{mh.1}) defined on an open 
chain with $\ell$ sites. Technically, we diagonalize $ H [ \Phi ]$ at $\tau = 0$, 
that is, $H_K$, with open boundary conditions on the single-mode wavefunction. 
A generic eigenmode, for $\Delta =w$, of $H_K$ with energy $E$ takes the form   

\beq
\Gamma_E = \sum_{ j = 1}^\ell \{ [ u_j ]^* c_j + [ v_j ]^* c_j^\dagger \}
\:\:\:\: , 
\label{oc.1}
\eneq
\noindent
with the wavefunctions $u_j , v_j$ solving the appropriate Bogoliubov-de Gennes (BdG)
equations obtained from the canonical commutation relation 
$ [ \Gamma_E , H_K ] = E \Gamma_E $. These are given by 

\begin{eqnarray}
 E u_j &=& - w \{ u_{ j + 1 } + u_{ j - 1 } \} - \mu u_j + w \{ v_{ j + 1 } - v_{ j - 1 } \}
 \nonumber \\
 E v_j &=& - w \{ u_{ j + 1} - u_{ j - 1 } \} + w \{ v_{ j + 1 }  + v_{ j - 1 } \} 
 + \mu v_j 
 \:\:\:\: ,
 \label{sg.1}
\end{eqnarray}
\noindent
for $1 < j < \ell$. At the endpoints ($j = 1 , \ell$), the equations 
take the form 

\begin{eqnarray}
 E u_1 &=& - w u_2 + w v_2 - \mu u_1 \nonumber \\
 E v_1 &=& - w u_2 + w v_2 + \mu v_1 
 \;\;\;\; , 
 \label{sg.2}
\end{eqnarray}
\noindent
and

\begin{eqnarray}
 E u_\ell &=& - w u_{ \ell - 1 } - w v_{ \ell - 1} - \mu u_\ell \nonumber \\
 E v_\ell &=& w u_{ \ell - 1} + w v_{ \ell - 1 } + \mu v_\ell 
 \:\:\:\: . 
 \label{sg.3}
\end{eqnarray}
\noindent
Requiring that Eqs.(\ref{sg.1}) are satisfied, the solution 
for $1 < j < \ell$ takes the form 

\beq
\left[ \begin{array}{c}
u_j \\ v_j         
       \end{array} \right] = e^{ i k j } \: \left[ \begin{array}{c}
u_k \\ v_k                                                     
                                                   \end{array} \right]
\:\:\:\: . 
\label{sg.4}
\eneq
\noindent
Imposing the wavefunctions in Eqs.(\ref{sg.4}) to be a solution of
Eqs.(\ref{sg.1}), one gets the system of equations in momentum 
space, given by 

\begin{eqnarray}
E u_k &=& - [ 2 w \cos ( k ) + \mu ] u_k + 2 i w \sin ( k ) v_k \nonumber \\
E v_k &=& - 2 i w \sin ( k ) u_k + [ 2 w \cos ( k ) + \mu ] v_k 
\:\:\:\: ,
\label{sg.5}
\end{eqnarray}
\noindent
supplemented with the boundary conditions 

\beq
u_0 + v_0 = u_{\ell + 1 } - v_{ \ell +1 } = 0 
\:\:\:\: . 
\label{sg.6}
\eneq
\noindent
From Eqs.(\ref{sg.5}), we obtain the dispersion relation (assuming $\mu > 0$)

\beq
E  = \pm \sqrt{ ( 2 w  - \mu )^2  + 8  w \mu \cos^2 \left( \frac{ k}{2} 
\right) }
\:\:\:\: . 
\label{sg.7}
\eneq
\noindent
Having defined the actual gap $\Delta_w$ as  $\Delta_w = | 2 w - \mu |$, Eq.(\ref{sg.7}) can be inverted, 
yielding the momentum of an excitation with energy $E $ as 

\beq
\cos \left( \frac{k}{2} \right) = \pm \sqrt{\frac{E^2 - \Delta_w^2}{8 w \mu}}
\:\:\:\: . 
\label{sg.8}
\eneq
\noindent
Solutions with energy $ | E | > \Delta_w$ correspond to real value of 
$k$. These can be readily written in a compact form, once one defines $\Psi$ 
so that

\begin{eqnarray}
 \cos ( \Psi ) &=& - \frac{2 w \cos ( k ) + \mu}{E} \nonumber \\
 \sin ( \Psi) &=& \frac{2 w \sin ( k ) }{E}
 \:\:\:\: . 
 \label{fc.5}
\end{eqnarray}
\noindent
Imposing the boundary conditions in Eq.(\ref{sg.6}), one 
eventually obtains the positive-energy solutions 

\beq
\left[ \begin{array}{c}
u_j \\ v_j         
       \end{array} \right]_+ = \sqrt{\frac{2}{\ell}} \: \left[ \begin{array}{c}
\cos \left( \frac{\Psi}{2} \right) \sin \left( k j + \frac{\Psi}{2} \right)  \\ 
-  \sin \left( \frac{\Psi}{2} \right)   \cos \left( k j + \frac{\Psi}{2} \right)                                     
                                    \end{array} \right]  
\:\:\:\:, 
\label{fc.6}
\eneq
\noindent
while the corresponding negative-energy solutions  are  recovered by acting with 
$\tau^x$ on the solution in Eq.(\ref{fc.6}), that is 

\beq
\left[ \begin{array}{c}
u_j \\ v_j         
       \end{array} \right]_- = \tau^x \left[ \begin{array}{c}
u_j \\ v_j                                               
                                             \end{array} \right]_+ = 
 \sqrt{\frac{2}{\ell}} \: \left[ \begin{array}{c}
-  \sin \left( \frac{\Psi}{2} \right)   \cos \left( k j + \frac{\Psi}{2} \right) 
 \\ 
\cos \left( \frac{\Psi}{2} \right) \sin \left( k j + \frac{\Psi}{2} \right)                                      
                                    \end{array} \right]  
\:\:\:\: . 
\label{fc.7}
\eneq
\noindent                   
The secular equation for the allowed values of $k$ is determined by the 
boundary condition at $j = \ell + 1$. It is given by 

\beq
\sin  [ k ( \ell + 1 ) + \Psi ] = 0 
\:\:\:\: . 
\label{ffc.1}
\eneq
\noindent
Sub-gap PFs are instead recovered for complex  values of $k$, which are
fixed by the condition  
 
\beq
\cos \left( \frac{k}{2} \right) = \pm  i \sqrt{\frac{\Delta_w^2 - E^2 }{8 w \mu}}
\:\:\:\: . 
\label{sg.9a}
\eneq
\noindent
To solve Eq.(\ref{sg.9a}), we now define the momentum for particle-like 
excitations as $k_p = \pi - i \delta $ and for hole-like excitations as 
$k_h = \pi + i \delta$, with 

\beq
\delta = 2 \sinh^{-1} \left\{ \sqrt{\frac{\Delta_w^2 - E^2 }{8 w \mu}}
\right\}
\:\:\:\: . 
\label{sg.10a}
\eneq
\noindent
As a result, the most general sub-gap eigenfunction with energy $E > 0 $ is given by 

\beq
\left[ \begin{array}{c}
u_j \\ v_j         
       \end{array} \right] = c \: (-1)^j \: \left[ \begin{array}{c}
 \cosh \left( \frac{\xi}{2} \right) \{ \alpha e^{ j \delta} + \beta e^{ - j \delta} \} \\
  \sinh \left( \frac{\xi}{2} \right) \{ \alpha e^{ j \delta}  - \beta e^{ - j \delta} \}
                                                   \end{array}  \right]
\;\;\;\;, 
\label{sg.15a}
\eneq
\noindent
with $\xi$ defined through the equations 

\begin{eqnarray}
 \cosh ( \xi ) &=& \frac{2 w \cosh ( \delta ) - \mu}{E} \nonumber \\
 \sinh ( \xi ) &=& \frac{2 w \sinh ( \delta ) }{E}
 \:\:\:\: ,
 \label{sg.14a}
\end{eqnarray}
\noindent
and the coefficients $\alpha$ and $\beta$ determined by the boundary conditions 
in Eqs.(\ref{sg.6}). Clearly, a state with energy $E > 0$  
comes together with the particle-hole conjugated one, with energy $- E$ \cite{kita}.  
In imposing the boundary  conditions in Eqs.(\ref{sg.6}), one finds that 
$\alpha e^{ \frac{\xi}{2}} + \beta e^{ - \frac{\xi}{2} } = 0$ and, more importantly, that
the allowed value of $E$ must satisfy the condition

\beq
\sinh [ \xi - ( \ell + 1 ) \delta ] = 0 \Rightarrow \xi ( E ) =  ( \ell + 1 ) \delta  ( E ) 
\:\:\:\: . 
\label{sg.16}
\eneq
\noindent
Eq.(\ref{sg.16}) is a transcendent equation, whose solution can in general only be derived numerically.
Yet, a simple approximate formula for the energy can be derived in the long chain limit, where one 
may assume that the energy is small enough to enable one to neglect  the dependence of $\delta$ on $E $ and, therefore, 
to make the approximation

\beq
\delta ( E ) \approx \delta_0 = 2 \sinh^{-1} \left\{ \sqrt{\frac{\Delta_w^2  }{8 w \mu}}
\right\}
\;\;\;\; .
\label{sg.19}
\eneq
\noindent
Thanks to Eq.(\ref{sg.19}), one therefore obtains 

\beq
E \sim \{ 2 w e^\delta - \mu \} e^{-\xi}  \approx \{ 2 w e^{ \delta_0 } - \mu \} e^{ - (\ell + 1 ) \delta_0 }
\:\:\:\: . 
\label{sg.20}
\eneq
\noindent
In general, even without knowing the explicit solution, one can identify  
the boundary of the  phase characterized by the low-lying modes by noting that, in order 
for Eq.(\ref{sg.16}) to be satisfied, $\xi ( E )$ must be real, which implies 
that $e^{ - \xi ( E )} > 0$. Therefore, one notes that 

\beq
e^{ - \xi ( E )} = \frac{2 w e^{ - \delta ( E )} - \mu}{E}
\;\;\;\; , 
\label{sg.21}
\eneq
\noindent
which implies that $\xi ( E )$ is real if, and only if, 

\beq
\frac{2 w}{\mu} > e^{ \delta ( E ) } = 
\frac{\sqrt{ ( 2 w + \mu )^2 - E^2} + \sqrt{( 2 w - \mu)^2 - E^2}
}{\sqrt{ ( 2 w + \mu )^2 - E^2} - \sqrt{( 2 w - \mu)^2 - E^2}}
\geq 1
\:\:\:\: . 
\label{sg.22}
\eneq
\noindent
Therefore, the phase characterized by the presence of low-lying PMFs  is defined by the condition 
$ \frac{2 w}{\mu} > 1$ \cite{kita}, even in the case of a finite-length chain.

\section{Excitation spectrum of the closed ring}
\label{closed_ring}

In this appendix, we derive the energy eigenvalues and the corresponding 
eigenmodes of $ H [ \Phi]$. Proceeding as for the open chain, for $1 < j < \ell$, the BdG equation take the same form as
in Eq.(\ref{sg.1}). At $j=1$, one gets

\begin{eqnarray}
 E u_1 &=& - w u_2 - \tau e^{ - \frac{i}{2} \Phi} u_\ell - \mu u_1 + w v_2 \nonumber \\
 E v_1 &=& - w u_2 + w v_2 + \tau e^{  \frac{i}{2} \Phi} v_\ell + \mu v_1 
 \:\:\:\: , 
 \label{sg.25}
\end{eqnarray}
\noindent
while at $j= \ell$, one obtains

\begin{eqnarray}
 E u_\ell &=& - w u_{ \ell - 1 } -  \tau e^{   \frac{i}{2} \Phi} u_1 - \mu u_\ell 
 - w v_{ \ell - 1 }  \nonumber \\
 E v_\ell &=& w u_{ \ell - 1 } + w v_{ \ell - 1 } + \tau    e^{ -  \frac{i}{2} \Phi} v_1 
 + \mu v_\ell
 \:\:\:\: . 
 \label{sg.26}
\end{eqnarray}
\noindent
In order for the conditions in Eqs.(\ref{sg.25},\ref{sg.26}) to be satisfied,
one has to modify the ansatz in  Eq.(\ref{sg.4}). For the sake of clarity, 
in the following we shall use  $\bar{u}_1 , \bar{v}_1$ 
and $\bar{u}_\ell , \bar{v}_\ell$ to respectively 
denote  the wavefunctions  at $j=1$ and at $j = \ell$. 
Therefore, using Eqs.(\ref{sg.25},\ref{sg.26}), 
one obtains 

\beq
\left[ \begin{array}{cc}
( E + \mu )    & \tau e^{ - \frac{i}{2} \Phi} \\      \tau e^{   \frac{i}{2} \Phi} 
& ( E + \mu ) 
       \end{array} \right] \left[ \begin{array}{c}
\bar{u}_1 \\ \bar{u}_\ell                                   
                                  \end{array} \right] = 
- w \left[ \begin{array}{c}
(u_2 - v_2 ) \\ (u_{ \ell - 1 } + v_{ \ell - 1} )             
           \end{array} \right] 
\;\;\;\; , 
\label{sg.27}
\eneq
\noindent
and 

\beq
\left[ \begin{array}{cc}
( E - \mu )    & - \tau e^{  \frac{i}{2} \Phi} \\     -  \tau e^{  -   \frac{i}{2} \Phi} 
& ( E - \mu ) 
       \end{array} \right] \left[ \begin{array}{c}
\bar{v}_1 \\ \bar{v}_\ell                                   
                                  \end{array} \right] = 
- w \left[ \begin{array}{c}
(u_2 - v_2 ) \\ - (u_{ \ell - 1 } + v_{ \ell - 1} )             
           \end{array} \right] 
\;\;\;\; . 
\label{sg.28}
\eneq
\noindent
On inverting Eqs.(\ref{sg.27}), one obtains 

\beq
 \left[ \begin{array}{c}
\bar{u}_1 \\ \bar{u}_\ell                                   
                                  \end{array} \right] = - \left\{ \frac{w}{( E + \mu )^2 - \tau^2} 
                                  \right\} 
\: \left[ \begin{array}{c}
\{  ( E + \mu ) ( u_2 - v_2 )     -   \tau e^{  - \frac{i}{2} \Phi} ( u_{ \ell - 1 } + v_{ \ell - 1 } ) \}         
\\
\{  - \tau e^{   \frac{i}{2} \Phi}  ( u_2 - v_2 ) +  ( E + \mu ) ( u_{ \ell - 1 } + v_{ \ell - 1 } ) \}  
          \end{array} \right] 
\;\;\;\; , 
\label{sg.29}
\eneq
\noindent
while, on inverting Eqs.(\ref{sg.28}), one rather gets

 \beq
 \left[ \begin{array}{c}
\bar{v}_1 \\ \bar{v}_\ell                                   
                                  \end{array} \right] = - \left\{ \frac{w}{( E - \mu )^2 - \tau^2} 
                                  \right\} 
\: \left[ \begin{array}{c}
\{  ( E - \mu ) ( u_2 - v_2 )     -   \tau e^{    \frac{i}{2} \Phi} ( u_{ \ell - 1 } + v_{ \ell - 1 } ) \}         
 \\
\{   \tau e^{ -    \frac{i}{2} \Phi}  ( u_2 - v_2 ) -   ( E - \mu ) ( u_{ \ell - 1 } + v_{ \ell - 1 } ) \}  
          \end{array} \right] 
\;\;\;\; .
\label{sg.30}
\eneq
\noindent  
Now, setting $j=2 , \ell - 1$, one obtains the BdG equations 

\begin{eqnarray}
 E u_2 &=& - w \{ u_3 + \bar{u}_1 \} - \mu u_2 + w \{ v_3 - \bar{v}_1 \} \nonumber \\
 E v_2 &=& - w \{ u_3 - \bar{u}_1 \} + w \{ v_3 + \bar{v}_1 \} + \mu v_2 
 \;\;\;\;\; , 
 \label{sg.31}
\end{eqnarray}
\noindent
and 

\begin{eqnarray}
 E u_{ \ell - 1  } &=& - w \{ u_{ \ell - 2} + \bar{u}_\ell \} - \mu u_{ \ell - 1 } + 
 w \{ \bar{v}_\ell - v_{ \ell - 2 } \} \nonumber \\
 E v_{\ell - 1} &=& - w \{ \bar{u}_\ell - u_{ \ell - 2} \} + w \{ \bar{v}_\ell + v_{ \ell - 2} \}
 + \mu v_{ \ell - 1 } 
 \:\:\:\: . 
 \label{sg.32}
\end{eqnarray}
\noindent
From Eqs.(\ref{sg.31},\ref{sg.32}) we see that, in order for the solution in 
Eq.(\ref{sg.15}) to hold for $1 < j < \ell$,  one must have 

\begin{eqnarray}
 \bar{u}_1 + \bar{v}_1 &=& u_1 + v_1 \nonumber \\
 \bar{u}_{ \ell } - \bar{v}_\ell &=& u_\ell - v_\ell 
 \:\:\:\: , 
 \label{sg.33}
\end{eqnarray}
\noindent
where, now, $u_1 , v_1 $ ($u_\ell , v_\ell$) denote the wavefunction 
$u_j , v_j$ evaluated at $j=1$ ($j= \ell $).  
Making a combined use of the above equations, one  
eventually gets to the  consistency conditions given by 

\begin{eqnarray}
 \bar{u}_1 + \bar{v}_1 &=& - w \left\{ \frac{E + \mu}{( E + \mu)^2 - \tau^2} + 
  \frac{E - \mu}{( E - \mu)^2 - \tau^2}  \right\} ( u_2 - v_2 ) + 
  \tau w \left\{ \frac{e^{ -\frac{i}{2} \Phi}}{( E + \mu)^2 - \tau^2} + 
  \frac{e^{ \frac{i}{2} \Phi}}{( E - \mu)^2 - \tau^2}  \right\} ( u_{ \ell - 1 } + 
  v_{ \ell -1 } ) \nonumber \\
   \bar{u}_\ell  - \bar{v}_\ell  &=& - w \left\{ \frac{E + \mu}{( E + \mu)^2 - \tau^2} + 
  \frac{E - \mu}{( E - \mu)^2 - \tau^2}  \right\} ( u_{ \ell - 1} +  v_{ \ell - 1}  ) + 
  \tau w \left\{ \frac{e^{ \frac{i}{2} \Phi}}{( E + \mu)^2 - \tau^2} + 
  \frac{e^{ - \frac{i}{2} \Phi}}{( E - \mu)^2 - \tau^2}  \right\} ( u_{ 2 } - 
  v_{ 2 } )
  \:\:\:\: . 
  \label{sg.34}
\end{eqnarray}
\noindent
We now make the ansatz that  a generic solution of energy $E$ takes  
the form 

\beq
\left[ \begin{array}{c}
        u_j \\ v_j 
       \end{array} \right]_+ = c \left[ \begin{array}{c}
\cos \left( \frac{\Psi}{2} \right) \{ a e^{ i k j } + b e^{ - i k j } \} \\
- i \sin \left( \frac{\Psi}{2} \right) \{ a e^{ i k j } - b e^{ - i k j } \} 
                                        \end{array} \right]
\:\:\:\: ,
\label{fc.8}
\eneq
\noindent
with $c$ being an appropriate normalization constant. 
Moreover, to simplify the notation, we set 

\begin{eqnarray}
 {\cal A} ( E ) &=& w \left\{ \frac{E + \mu}{( E + \mu)^2 - \tau^2} + 
  \frac{E - \mu}{( E - \mu)^2 - \tau^2}  \right\} \nonumber \\
{\cal B} ( E ; \Phi ) &=&  \tau w \left\{ \frac{e^{- \frac{i}{2} \Phi}}{( E + \mu)^2 - \tau^2} + 
  \frac{e^{ \frac{i}{2} \Phi}}{( E - \mu)^2 - \tau^2}  \right\} 
  \;\;\;\; . 
  \label{sg.36}
\end{eqnarray}
\noindent
Therefore, we obtain the system of algebraic equations for $a$ and $b$ given by 

\begin{eqnarray}
 && \{ e^{ i k - i \frac{\Psi}{2} } + {\cal A} ( E ) e^{ 2 i k + i \frac{\Psi}{2} } - {\cal B} [ E ; \Phi ] 
 e^{ i k ( \ell - 1 ) - i \frac{\Psi}{2} } \} a + \{ e^{ - i k + i \frac{\Psi}{2} } + {\cal A} ( E ) 
 e^{ - 2 i k -  i \frac{\Psi}{2} } - {\cal B} [ E ; \Phi ] 
 e^{ - i k ( \ell - 1 ) + i \frac{\Psi}{2} } \} b = 0 \nonumber \\
  && \{ e^{ i k \ell +  i \frac{\Psi}{2} } + {\cal A} ( E ) e^{  i k ( \ell - 1 )  - i \frac{\Psi}{2} } - {\cal B}^* [ E ; \Phi ] 
 e^{ 2 i k  + i \frac{\Psi}{2} } \} a + \{ e^{ - i k  \ell -  i \frac{\Psi}{2} } + {\cal A} ( E ) 
 e^{ - i k ( \ell - 1 )  +  i \frac{\Psi}{2} } - {\cal B}^* [ E ; \Phi ] 
 e^{ - 2 i k -  i \frac{\Psi}{2} } \} b = 0
 \:\:\:\: . 
 \label{fc.10}
\end{eqnarray}
\noindent
On requiring Eqs.(\ref{fc.10}) to provide nontrivial solutions for 
$a$ and $b$, one readily obtains the secular equation for the allowed values of 
the energy $| E | > \Delta_w$ in the form 

\begin{eqnarray}
 && - \sin [ k ( \ell - 1 ) +  \Psi ] - ( {\cal A}^2 ( E ) - | {\cal B} [ E ; \Phi ] |^2 )
\sin  [ k ( \ell - 3 ) - \Psi ] \nonumber \\
&& - 2 {\cal A} ( E ) \sin [ k ( \ell - 2 ) ] + 2 \Re e {\cal B} [ E ; \Phi ] 
\sin [ k + \Psi ] = 0 
\:\:\:\: . 
\label{fc.11}
\end{eqnarray}
\noindent
Next, we consider the  equation for  subgap PMF energies. These correspond 
to complex values of $k$ satisfying 
 
\beq
\cos \left( \frac{k}{2} \right) = \pm  i \sqrt{\frac{\Delta_w^2 - E^2 }{8 w \mu}}
\:\:\:\: . 
\label{sg.9}
\eneq
\noindent
To solve Eq.(\ref{sg.9}), we now define the momentum for particle-like 
excitations as $k_p = \pi - i \delta $ and for hole-like excitations as 
$k_h = \pi + i \delta$, with 

\beq
\delta = 2 \sinh^{-1} \left\{ \sqrt{\frac{\Delta_w^2 - E^2 }{8 w \mu}}
\right\}
\:\:\:\: . 
\label{sg.10}
\eneq
\noindent
As a result, one finds that the positive-energy PMF wavefunction   is given by 

\beq
\left[ \begin{array}{c}
u_j \\ v_j         
       \end{array} \right] = c \: (-1)^j \: \left[ \begin{array}{c}
 \cosh \left( \frac{\xi}{2} \right) \{ \alpha e^{ j \delta} + \beta e^{ - j \delta} \} \\
  \sinh \left( \frac{\xi}{2} \right) \{ \alpha e^{ j \delta}  - \beta e^{ - j \delta} \}
                                                   \end{array}  \right]
\;\;\;\;, 
\label{sg.15}
\eneq
\noindent
with $\xi$ defined through the equations 

\begin{eqnarray}
 \cosh ( \xi ) &=& \frac{2 w \cosh ( \delta ) - \mu}{E} \nonumber \\
 \sinh ( \xi ) &=& \frac{2 w \sinh ( \delta ) }{E}
 \:\:\:\: ,
 \label{sg.14}
\end{eqnarray}
\noindent
and the coefficients $\alpha$ and $\beta$ determined by the appropriate boundary 
conditions for the allowed wavefunctions. We therefore trade Eqs.(\ref{sg.34}) for  
the following system in the unknowns $\alpha , \beta$:

\begin{eqnarray}
&& \{ e^{ \delta + \frac{\xi}{2} } - e^{ 2  \delta - \frac{\xi}{2} } {\cal A} ( E ) - 
 e^{ ( \ell - 1 ) \delta + \frac{\xi}{2} } {\cal B} ( E ; \Phi) \} \alpha + 
  \{ e^{ -\delta - \frac{\xi}{2} } - e^{ -2  \delta + \frac{\xi}{2} } {\cal A} ( E ) - 
 e^{- ( \ell - 1 ) \delta - \frac{\xi}{2} } {\cal B} ( E ; \Phi) \} \beta = 0 \nonumber \\
&&  \{ e^{ \ell \delta -  \frac{\xi}{2} } - e^{ (\ell - 1 )   \delta +
\frac{\xi}{2} } {\cal A} ( E ) - 
 e^{ 2 \delta - \frac{\xi}{2} } {\cal B}^* ( E ; \Phi) \} \alpha + 
  \{ e^{ -\ell \delta +  \frac{\xi}{2} } - e^{ -(\ell - 1 )   \delta -
\frac{\xi}{2} } {\cal A} ( E ) - 
 e^{ - 2 \delta +  \frac{\xi}{2} } {\cal B}^* ( E ; \Phi) \} \beta = 0 
 \:\:\:\: .
 \label{sg.37}
\end{eqnarray}
\noindent
The system in Eq.(\ref{sg.37}) admits a nontrivial solution for $\alpha$ and $\beta$ only provided that 
 the following secular equation for the energy eigenvalue $E$ is satisfied

\beq
- \sinh [ ( \ell - 1 ) \delta    - \xi   ] + 2 {\cal A} ( E ) \sinh [ ( \ell - 2 ) \delta  ] +
2 \Re e ( {\cal B}  ( E ; \Phi ) ) 
\sinh [ \delta    - \xi ] - \{ {\cal A}^2 ( E ) - | {\cal B}  ( E ; \Phi ) |^2 \} 
\sinh [ ( \ell - 3 ) \delta  + \xi ]  = 0 
\:\:\:\: . 
\label{sg.38}
\eneq
\noindent
Clearly, Eqs.(\ref{sg.37},\ref{sg.38}) are consistent with the solution for 
$\tau  = 0$. Indeed, as $\tau = 0 $ (open chain limit), one obtains 
that (apart for a constant) $\alpha = e^{ - \frac{\xi}{2}}$ and 
$\beta = - e^{ \frac{\xi}{2}}$.  Also, we obtain that ${\cal B} ( E ; \Phi ) = 0$ and, 
as a result, Eqs.(\ref{sg.37}) take the form 

\begin{eqnarray}
 && \{ e^{ \delta  + \frac{\xi}{2} } - e^{ 2  \delta - \frac{\xi}{2} } {\cal A} ( E )   \} \alpha + 
  \{ e^{ -\delta - \frac{\xi}{2} } - e^{ -2  \delta + \frac{\xi}{2} } {\cal A} (E  )  \} \beta = 0 \nonumber \\
&&  \{ e^{ \ell \delta -  \frac{\xi}{2} } - e^{ (\ell - 1 )   \delta +
\frac{\xi}{2} } {\cal A} ( E )   \} \alpha + 
  \{ e^{ -\ell \delta +  \frac{\xi}{2} } - e^{ -(\ell - 1 )   \delta -
\frac{\xi}{2} } {\cal A} ( E  )   \} \beta = 0 
\:\:\:\: . 
\label{sg.39}
\end{eqnarray}
\noindent
Using the explicit expression for ${\cal A} ( E )$, Eqs.(\ref{sg.39}) yield 

\begin{eqnarray}
 && \alpha e^{ \delta + \frac{\xi}{2} }  + \beta  e^{ -\delta - \frac{\xi}{2} }
 - \left[ \frac{w}{E  + \mu} + \frac{w}{E  - \mu} \right] \{  \alpha  e^{ 2  \delta - \frac{\xi}{2} } 
 +  \beta  e^{ -2  \delta + \frac{\xi}{2} } \} = 0 \nonumber \\
 \nonumber \\
   && \alpha  e^{ \ell \delta -  \frac{\xi}{2} }+ \beta e^{ -\ell \delta +  \frac{\xi}{2} } 
   - \left[ \frac{w}{E  + \mu} + \frac{w}{E  - \mu} \right] \{ \alpha e^{ (\ell - 1 )   \delta +
\frac{\xi}{2} }  + 
   \beta  e^{ -(\ell - 1 )   \delta -
\frac{\xi}{2} } \} = 0 
\:\:\:\:,
\label{sg.40}
\end{eqnarray}
\noindent
that is 

\begin{eqnarray}
 && u_1 + v_1  + \left[ \frac{w}{E  + \mu} + \frac{w}{E - \mu} \right] \{ u_2 - v_2 \} = 0 \nonumber \\
 && v_\ell - v_\ell + \left[ \frac{w}{E  + \mu} + \frac{w}{E - \mu} \right] \{ u_{ \ell - 1 } + v_{\ell - 1 }  \} = 0 
 \:\:\:\: . 
 \label{sg.41}
\end{eqnarray}
\noindent
We now  see that Eqs.(\ref{sg.41}) respectively 
imply $u_0 + v_0 = 0$, and $u_{ \ell + 1 } - v_{ \ell + 1 } =0$, which are the appropriate equations 
for $\tau= 0 $. Of course, in general  Eq.(\ref{fc.11},\ref{sg.38}) for the PMF energy 
appear to be quite formidable 
and one  has to numerically solve them to recover the   functional dependence of the energy $E  
$ on $\Phi$, just as we do in the main text. Nevertheless, one may effectively use 
Eq.(\ref{sg.38}) to recover under which conditions, and at which values of $\Phi = \Phi_*$, 
the PMFs energy levels may cross each other, which is a key point in the application
of our technique.  In order to recover this point,   we 
 consider the zero-energy limit of Eq.(\ref{sg.38}), thus  obtaining 

\beq
{\cal B}^2 [ 0 ; \Phi_* ] - 2 {\cal B} [ 0 ; \Phi_* ] e^{  - ( \ell - 2 ) \delta_0} + e^{ - 2 ( \ell - 2 ) \delta_0}  = 0 
\;\;\;\; , 
\label{sg.46}
\eneq
\noindent
where we have taken into account that ${\cal B} [ 0 ; \Phi] $ is real. Eq.(\ref{sg.46})
is then solved by setting 

\beq
\frac{2 w \tau}{\mu^2 - \tau^2 } \cos \left ( \frac{\Phi_*}{2} \right) =   e^{  - ( \ell - 2 ) \delta_0} \Rightarrow
\cos \left( \frac{\Phi_*}{2} \right) = \frac{\mu^2 - \tau^2}{2 w \tau }  e^{  - ( \ell - 2 ) \delta_0}
\;\;\;\; . 
\label{sg.47}
\eneq
\noindent
 Eq.(\ref{sg.47}) admits real solutions for $\Phi_*$ only provided that 
$ \left| \frac{\mu^2 - \tau^2}{2 w \tau }  e^{  - ( \ell - 2 ) \delta_0} \right| \leq 1$, which is 
always obeyed for either a long enough chain, or a large enough $\tau$ (or both). A detailed 
discussion about this point is provided in the main text. Here, we point out an important consequence 
of Eq.(\ref{sg.47}), namely, even for a finite-size ring, it is still possible to recover two
zero-energy MMs, provided $\Phi$ is properly tuned. To see this, at a given $\Phi$, we define the 
mode operators corresponding to the PMFs, $\Gamma_0 [ \Phi ]  , \Gamma_0^\dagger [ \Phi ] $, respectively given by 

\begin{eqnarray}
 \Gamma_0 [ \Phi ] &=& \sum_{ j = 1}^\ell \{ [ u_j ]^* c_j + [ v_j ]^* c_j^\dagger \}
 \nonumber \\
  \Gamma_0^\dagger [ \Phi ] &=& \sum_{ j = 1}^\ell \{  u_j  c_j^\dagger  +  v_j  c_j \}
  \:\:\:\: , 
  \label{ssg.1}
\end{eqnarray}
\noindent
with $u_j , v_j$ being the wavefunctions in Eq.(\ref{sg.15}). By definition, we then obtain 

\begin{eqnarray}
[ \Gamma_0 [ \Phi ] , H ] &=& \epsilon_0 [ \Phi ]  \Gamma_0 [ \Phi ]  \nonumber \\
[ \Gamma_0^\dagger [ \Phi ] , H ] &=& - \epsilon_0 [ \Phi ] \Gamma_0^\dagger [ \Phi ] 
\:\:\:\: , 
\label{ssg.2}
\end{eqnarray}
\noindent
with $\epsilon_0 [ \Phi ] $ being the solution of Eq.(\ref{sg.38}) at a given $\Phi$. Thus, 
for $\Phi = \Phi_*$, one obtains 

\beq
[ \Gamma_{ 0 , *} , H ] = [ \Gamma_{ 0 , *}^\dagger , H ] = 0 
 \;\;\;\; 
\label{ssg.3}
\eneq
\noindent
with $\Gamma_{ 0 , * } = \Gamma_0 [ \Phi = \Phi_* ]$. As a result, for $\Phi = \Phi_*$ one 
may define the zero-mode MMs $\gamma_1 , \gamma_2$ by simply setting 

\begin{eqnarray}
 \gamma_1 &=& \Gamma_{ 0 , *} + \Gamma_{ 0 , *}^\dagger \nonumber \\
 \gamma_2 &=& - i \{ \Gamma_{ 0 , * } - \Gamma_{ 0 , *}^\dagger \}
 \:\:\:\: . 
 \label{ssg.4}
\end{eqnarray}
\noindent
As we discuss in the paper, the existence of $\gamma_1 , \gamma_2$ as zero-energy MMs is 
quite robust against the effects of disorder. For comparison, in the main text of the 
paper we also consider the dependence on $\Phi$ of the subgap states in a superconducting ring made with 
an s-wave one-dimensional superconductor closed with a weak link of strength $\tau$. This is derived from 
the corresponding  solution of the Bogoliubov-de Gennes equations for the wavefunction of a spin-$\sigma$ 
quasiparticle, $u_{ j } , v_j$ which, for the s-wave ring of length $\ell$ and for the superconducting 
gap $\Delta = w$ (in analogy to the simplifying limit we take above for the p-wave superconductor), are given by 
\cite{ngcg}

\begin{eqnarray}
 E u_{ j  } &=& - w \{ u_{ j + 1 } + u_{ j - 1 } \} - \mu u_j + \sigma w v_j \nonumber \\
E v_j &=& \sigma w v_j + w \{ v_{j+1} + v_{ j - 1 } \} + \mu v_j  
\:\:\:\: ,
\label{sw.1}
\end{eqnarray}
\noindent
for $1 < j < \ell$, and by 

\begin{eqnarray}
 E u_1 &=& - w u_2 - \tau e^{ - \frac{i}{2} \Phi} u_\ell - \mu u_1 + \sigma w v_1 \nonumber \\
 E v_1 &=& \sigma w u_1 + w v_2 + \tau e^{ \frac{i}{2} \Phi} v_\ell + \mu v_1 
 \;\;\;\; , 
 \label{sw.2}
\end{eqnarray}
\noindent
for $j=1$ and, finally, by 

\begin{eqnarray}
 E u_\ell &=& - w u_{\ell - 1 } -   \tau e^{ \frac{i}{2} \Phi} u_1 - \mu u_\ell + \sigma w v_\ell \nonumber \\
 E v_\ell &=& \sigma w u_\ell + w v_{\ell -1 } + \tau e^{ - \frac{i}{2} \Phi} v_1 + \mu v_\ell 
 \;\;\;\; , 
 \label{sw.3}
\end{eqnarray}
\noindent
for $j = \ell$. Eq.(\ref{sw.1}) is solved by a wavefunction of the form 

\beq
\left[ \begin{array}{c}
u_j \\ v_j         
       \end{array} \right] = \left[ \begin{array}{c}
                                     u \\ v 
                                    \end{array} \right] e^{ i k j } 
\;\;\;\; , 
\label{sw.4}
\eneq
\noindent
provided $u , v$ satisfy the secular equation

\begin{eqnarray}
 E u &=& - \{ 2 w \cos ( k ) + \mu \} u + \sigma w v \nonumber \\
 E v &=& \sigma w u + \{ 2 w \cos ( k ) + \mu \} v
 \;\;\;\; . 
 \label{sw.5}
\end{eqnarray}
\noindent
Imposing the wavefunction in Eq.(\ref{sw.5}) to satisfy Eqs.(\ref{sw.2},\ref{sw.3}) as well 
implies the interface conditions across the weak link given by

\begin{eqnarray}
 0 &=& w u_0 - \tau e^{ - \frac{i}{2} \Phi} u_\ell \nonumber \\
 0 &=& w v_0 - \tau e^{   \frac{i}{2} \Phi} v_\ell \nonumber \\
 0 &=& \tau e^{ \frac{i}{2} \Phi} u_1 - w u_{ \ell + 1 } \nonumber \\
  0 &=& \tau e^{ - \frac{i}{2} \Phi} v_1 - w v_{ \ell + 1 }
  \:\:\:\: . 
  \label{sw.6}
 \end{eqnarray}
\noindent
Let us, now, look for subgap solutions, with energy $| E | < w$.  In this case, we 
obtain that $ k = \pm \left[ \frac{\pi}{2} \pm i q \right]$, and that, accordingly, 
the generic subgap solution is given by

\beq
\left[ \begin{array}{c}
        u_j \\ v_j 
       \end{array} \right] = 
C \: \left[ \begin{array}{c}
 ( i^j e^{ - q j } \alpha_{ p , + } + i^{-j} e^{ q j } \alpha_{ p , - } ) e^{ \frac{i}{2} \xi } 
 +  ( i^{-j} e^{ - q j } \alpha_{ h , + } + i^{j} e^{ q j } \alpha_{ h , - } ) 
 e^{ - \frac{i}{2} \xi }  \\
  ( i^j e^{ - q j } \alpha_{ p , + } + i^{-j} e^{ q j } \alpha_{ p , - } ) e^{ - \frac{i}{2} \xi } 
 +  ( i^{-j} e^{ - q j } \alpha_{ h , + } + i^{j} e^{ q j } \alpha_{ h , - } ) 
 e^{  \frac{i}{2} \xi } 
            \end{array} \right]
\;\;\;\; , 
\label{s.wave1}
\eneq
\noindent
with 

\begin{eqnarray}
 \sinh [ q ( E ) ] &=& \sqrt{\frac{w^2 - E^2}{4 w^2}} \nonumber \\
 \tan [ \xi ( E ) ]  &=&  \sqrt{\frac{w^2 - E^2}{E^2}} 
 \:\:\:\: ,
 \label{sw.1b}
\end{eqnarray}
\noindent
and $\alpha_{ p , + } , \alpha_{ p , - } , \alpha_{ h , + } , \alpha_{ h , - }$ coefficients. 
On requiring to recover a nontrivial solution for the above coefficients, one eventually finds the secular
equation for the allowed energy eigenvalues, which is given by 

\beq
4 w \tau e^{ \ell q } ( e^{ 2 q } - 1 ) \cos \left( \frac{\Phi}{2} \right) 
=  4 w^2 \{ e^{ 2 q ( \ell + 1 ) } - 1 + \tau^2 ( e^{ 2 q } - e^{ 2 q \ell } ) \} 
\:\:\:\: . 
\label{sw.x1}
\eneq
\noindent
Eq.(\ref{sw.x1}) has been used to work out the subgap energy levels in the s-wave case, which 
have  been discussed in the main text in comparison with the ones in the p-wave case.

\bibliography{disorder_biblio_revised}

\end{document}